# A review of the literature on citation impact indicators

Ludo Waltman

Centre for Science and Technology Studies, Leiden University, The Netherlands
waltmanlr@cwts.leidenuniv.nl

Citation impact indicators nowadays play an important role in research evaluation, and consequently these indicators have received a lot of attention in the bibliometric and scientometric literature. This paper provides an in-depth review of the literature on citation impact indicators. First, an overview is given of the literature on bibliographic databases that can be used to calculate citation impact indicators (Web of Science, Scopus, and Google Scholar). Next, selected topics in the literature on citation impact indicators are reviewed in detail. The first topic is the selection of publications and citations to be included in the calculation of citation impact indicators. The second topic is the normalization of citation impact indicators, in particular normalization for field differences. Counting methods for dealing with co-authored publications are the third topic, and citation impact indicators for journals are the last topic. The paper concludes by offering some recommendations for future research.

## 1. Introduction

Citation impact indicators are indicators of scientific impact that are based on an analysis of the citations received by scientific publications. Citation impact indicators may provide information on the impact of individual publications, but more often they provide information on the impact of research units such as researchers, research groups, research institutions, countries, or journals. In that case, citation impact indicators are based on an analysis of the citations received by the entire publication oeuvre of a research unit. Well-known examples of citation impact indicators are the journal impact factor (Garfield, 1972) and the h-index (Hirsch, 2005).

Citation impact indicators nowadays play a prominent role in the evaluation of scientific research. The importance of citation impact indicators in the context of research evaluation has increased a lot during the past decades, and this is reflected in a rapidly growing body of scientific literature in which citation impact indicators are studied. Most of this literature can be found in journals in the fields of bibliometrics, scientometrics, and research evaluation, although contributions to this literature are also often made by researchers from other fields.

In this paper, I present an in-depth review of the literature on citation impact indicators. This review aims to serve both researchers studying citation impact indicators and practitioners working with these indicators. An overview is provided of different citation impact indicators that have been proposed in the literature and, more generally, of different choices that can be made in the construction of citation impact indicators. In practice, citation impact indicators are calculated based on data obtained from bibliographic databases. The literature on bibliographic databases is therefore reviewed as well, focusing on the three most popular multidisciplinary databases: Web of Science (WoS), Scopus, and Google Scholar.

The literature on citation impact indicators is rather large, and it is not possible to cover the entire literature in this review. Because of this, there are various topics related to citation impact indicators that are not discussed in this review. First of all, no detailed review of the literature on the h-index and related indicators is provided. During recent years, a large literature on this topic has emerged, but reviews of this literature can already be found elsewhere (Alonso et al., 2009; Egghe, 2010; Norris & Oppenheim, 2010; Panaretos & Malesios, 2009). There also is a literature in which citation impact indicators are studied from a purely mathematical point of view. This literature is of less interest to practitioners working with citation impact indicators, and therefore I have chosen not to include it in this review. Furthermore, this review also does not cover literature on the interpretation of citation impact indicators (Bornmann & Daniel, 2008; Nicolaisen, 2007), literature on the practical application of citation impact indicators in the context of research evaluation, and literature on the correlation between citation impact indicators and peer review. I refer to Moed (2005) for an introduction into these topics. Finally, no discussion of the historical development of the literature on citation impact indicators is provided. Such a historical account is offered by De Bellis (2009).

This paper presents the first large-scale review of the literature on citation impact indicators. However, there is some related work to which I would like to draw attention. Vinkler (2010) offers a systematic overview of scientometric indicators for research evaluation. This overview has a broader scope than the literature review provided in the present paper, but its coverage of the recent literature on citation impact indicators is less extensive. Wildgaard et al. (2014) present a review of the literature on bibliometric indicators for assessing the performance of individual researchers. A limitation of this review is that it focuses exclusively on individual researchers and does not consider other research units. Finally, Mingers and Leydesdorff (2015) provide a review of the entire scientometric literature. This review has a broad scope and citation impact indicators are just one topic covered in the review.

An earlier version of the literature review presented in this paper appeared in a report prepared for the Higher Education Funding Council for England (HEFCE; Wouters et al., 2015). This report provides an overview of the literature on the following four topics: (1) citation impact indicators, (2) effects of the use of indicators in research evaluation, (3) relation between indicators and peer review, and (4) alternative indicators for research evaluation. The reviews on topics (2) and (4) have also been published separately (De Rijcke et al., 2015; Kousha & Thelwall, 2015; Thelwall & Kousha, 2015a, 2015b).

The organization of this paper is as follows. First, the methodology used to collect the literature included in this review is discussed in Section 2. Next, a review of the literature on bibliographic databases is provided in Section 3. An overview of the most basic citation impact indicators is then presented in Section 4. Based on this overview, selected topics in the literature on citation impact indicators are reviewed in Sections 5 to 8. Section 5 deals with the selection of publications and citations to be included in the calculation of citation impact indicators. Sections 6, 7, and 8 cover the topics of normalization, counting methods, and citation impact indicators for journals, respectively. Finally, some recommendations for future research are made in Section 9.

## 2. Methodology

I collected the literature included in this review using a semi-systematic methodology. First, based on my prior knowledge of the literature, I chose the topics to be covered in the review. Next, for each topic, I selected an initial set of relevant publications. This was again done based on my prior knowledge of the literature, but in addition some preliminary literature searches were performed as well. Given a set of relevant publications on a certain topic, additional relevant publications were identified in a systematic way.

To systematically identify relevant publications on a given topic, I used CitNetExplorer (Van Eck & Waltman, 2014a, 2014b). CitNetExplorer is a software tool for visualizing and analyzing citation networks of scientific publications. The tool is freely available at www.citnetexplorer.nl. I first downloaded data from the WoS database on about 26,000 publications that appeared in the 14 journals listed in Table 1. I selected these journals either because they are core journals in the fields of bibliometrics, scientometrics, and research evaluation or because they have strong citation relations to core journals in these fields. I then provided the data on the 26,000 publications as input to CitNetExplorer. CitNetExplorer then constructed the citation network of the 26,000 publications. In addition, CitNetExplorer identified publications (not only journal publications but also for instance books) that did not appear in the 14 journals listed in Table 1 but that were cited at least ten times in these journals. These publications were also included in the citation network. In this way, a citation network of almost 30,000 publications was obtained.

Table 1. Journals included in the systematic literature search.

| | |
|---|---|
| American Documentation | Journal of Informetrics |
| Annual Review of Information Science and Technology | Journal of the American Society for Information Science |
| ASLIB Proceedings | Journal of the American Society for Information Science and Technology |
| Information Processing and Management | |
| Information Scientist | Research Evaluation |
| Information Storage and Retrieval | Research Policy |
| Journal of Documentation | Scientometrics |
| Journal of Information Science | |

Given an initial set of relevant publications on a certain topic, I used CitNetExplorer to identify additional publications that could potentially be of relevance. This was done based on citation relations between publications. For instance, all publications cited by or citing to publications already classified as relevant were identified. Or alternatively, publications with at least a certain minimum number of citation relations (e.g., three or four citation relations) with publications already classified as relevant were identified. For each publication identified by CitNetExplorer, I then manually determined (e.g., based on the title and abstract of the publication) whether the publication is indeed of relevance to the topic of interest or not. In this way, the set of relevant publications was extended. The above steps could then be repeated in order to identify additional relevant publications. A number of iterations were usually performed until all relevant publications on a certain topic seemed to have been found.

Two further comments should be made on the way in which the publications included in this review were selected. First, it should be emphasized that the primary aim of this review is to provide an overview of the current state of the art in the literature on citation impact indicators. The focus of the review therefore is mainly on the more recent literature on citation impact indicators. No special attention is paid to the historical development of the literature. Second, it should be mentioned that this review does not provide an exhaustive overview of the literature on citation impact indicators. Given the size of the literature, providing an exhaustive overview in which all relevant publications are included is hardly possible. As already pointed out in Section 1, this review focuses on selected topics studied in the literature on citation impact indicators. However, even the literature on these selected topics cannot be covered in a fully comprehensive way. This review therefore includes the publications that were considered to be most relevant or most interesting. In some cases, citation counts were used to support decisions on which publications to include in the review and which publications to exclude.

## 3. Bibliographic databases

The three most important databases available for performing citation analyses are WoS, Scopus, and Google Scholar. There are more databases available, but these usually cover only a limited number of scientific fields. Moreover, some of these databases do not contain data on the references of publications, and these databases therefore cannot be used to calculate citation impact indicators. I start by summarizing some key features of WoS, Scopus, and Google Scholar.

WoS is a subscription-based database that comprises a number of citation indices. The best-known citation indices are the Science Citation Index Expanded, the Social Sciences Citation Index, and the Arts & Humanities Citation Index. These citation indices cover journals and book series. Nowadays, WoS also offers a Conference Proceedings Citation Index and a Book Citation Index, covering conference proceedings and books. Recently, the Emerging Sources Citation Index was added to WoS. This citation index aims to cover scientific literature of regional importance and in emerging scientific fields. WoS was originally owned by the Institute for Scientific Information. Its current owner is Thomson Reuters. The use of WoS for performing citation analyses has a long history, and therefore the characteristics of this database have been studied in significant detail. For instance, Moed (2005, Chapter 7) and Larsen and Von Ins (2010) analyze the coverage of WoS, García-Pérez (2011) draws attention to the issue of incorrect citation relations in WoS, Michels and Schmoch (2012) investigate the growth of WoS, Harzing (2013a) studies the document type classification of WoS, Olensky et al. (in press) analyze the accuracy of the citation matching algorithm of WoS, and Zhang et al. (in press) perform a comparison between author keywords and algorithmically selected keywords ('KeyWords Plus') in WoS.

Like WoS, Scopus is a subscription-based database. In addition to journals, Scopus also covers trade publications, book series, conference proceedings, and books. Scopus is owned by Elsevier and was launched in 2004. The characteristics of the Scopus database have been studied less extensively than those of WoS, but some work has been done. In particular, Franceschini et al. (2015a) discuss the assignment of incorrect DOIs to publications in Scopus, Kawashima and Tomizawa (2015) study the accuracy of Scopus author identifiers, and Valderrama-Zurián et al. (2015) analyze the problem of duplicate publications in Scopus. A more general critical discussion on the data quality of Scopus is provided by Franceschini et al. (2016). Furthermore, a combined analysis of WoS and Scopus relating to the problem of missing citation relations is reported by Franceschini et al. (2013, 2014, 2015b, in press), while a comparative analysis of the accuracy of the journal classification systems of WoS and Scopus is presented by Wang and Waltman (2015). Gorraiz et al. (2016) study the availability of DOIs in WoS and Scopus.

Google Scholar was also launched in 2004. It indexes scholarly literature that is available online on the web. This includes not only publications in journals and conference proceedings, but also for instance books, theses, preprints, and technical reports. Google Scholar is made freely available by Google. It should be emphasized that Google Scholar is of a very different nature than WoS and Scopus. It is primarily a search engine for scholarly literature, and it provides only very limited bibliographic meta data on publications. Little is known about the coverage of Google Scholar. For instance, there is no list available of sources that are covered by Google Scholar. In a recent study by Khabsa and Giles (2014), it is estimated that Google Scholar indexes about 100 million English-language documents, representing almost 87% of all English-language scholarly documents available on the web. Another study by Orduna-Malea et al. (2015) estimates that the total number of documents indexed by Google Scholar, without any language restriction, is between 160 and 165 million.

Most institutions with a subscription to WoS or Scopus have access to these databases through a web interface. The WoS and Scopus web interfaces can be used for performing simple citation analyses at a relatively small scale. Advanced citation analyses at a larger scale require direct access to the full WoS or Scopus database, without the restrictions imposed by a web interface. Professional bibliometric centers often have direct access to the full WoS or Scopus database. An alternative way of performing advanced citation analyses is the use of specialized web-based tools such as InCites and SciVal. InCites is provided by Thomson Reuters based on WoS, and SciVal is provided by Elsevier based on Scopus. Performing large-scale citation analyses using Google Scholar is more difficult, because the only way to access Google Scholar is through its web interface. It is not possible to get direct access to the full Google Scholar database. Citation analyses based on Google Scholar are sometimes performed using a software tool called Publish or Perish (Harzing, 2010).

Below, an overview of the literature on WoS, Scopus, and Google Scholar is provided, focusing in particular on studies of the coverage of the different databases. It should be emphasized that all three databases are in continuous development (e.g., Chen, 2010; De Winter et al., 2014; Harzing, 2013b, 2014; Michels & Schmoch, 2012). Results reported in the literature, especially in less recent work, may therefore not be up-to-date anymore.

**3.1. Comparing Web of Science and Scopus**

The most comprehensive comparison between WoS and Scopus is reported by Visser and Moed (2008). By matching publications in Scopus with publications in WoS, they establish that 97% of all publications from 2005 covered by WoS are also covered by Scopus. Hence, WoS can almost be considered a perfect subset of Scopus. Looking at publications submitted to the 2001 Research Assessment Exercise, Scopus coverage turns out to be broader than WoS coverage especially in the subject group 'Subjects Allied to Health' and to a lesser degree in the subject group 'Engineering & Computer Science'. Visser and Moed (2008) note that when Scopus is used in a citation analysis it may be preferable to work with a subset of all Scopus data instead of the entire database. As reported by López-Illescas et al. (2008, 2009) based on an analysis in the field of oncology, journals covered by Scopus and not covered by WoS tend to have a low citation impact and tend to be more nationally oriented. Including these journals in a citation analysis may significantly reduce the average citation impact of certain countries.

The observation that Scopus has a broader coverage than WoS is made in various other studies as well. In an analysis of researchers in the field of human-computer interaction, Meho and Rogers (2008) observe that Scopus has a broader coverage of conference proceedings than WoS. Gavel and Iselid (2008) find that Scopus has a broader journal coverage than WoS especially in science, technology, and medicine. Norris and Oppenheim (2007) observe that Scopus has a broader coverage than WoS of social science publications submitted to the 2001 Research Assessment Exercise. In an analysis of Slovenian publications, Bartol et al. (2014) report that Scopus has a broader coverage than WoS in the social sciences, humanities, and engineering & technology. In a study of publications of two Portuguese universities, Vieira and Gomes (2009) observe publications that are covered by Scopus and not by WoS, but the reverse situation is found as well. Likewise, Cavacini (2015) reports that Scopus has a broader coverage of journal publications in computer science than WoS, although each of the two databases covers some publications that are not indexed in the other database. In a comprehensive analysis of the coverage of WoS and Scopus at the level of journals, Mongeon and Paul-Hus (2016) find that Scopus covers a much larger number of journals than WoS and that almost all journals covered by WoS are also covered by Scopus. Mongeon and Paul-Hus (2016) emphasize that both in WoS and in Scopus social sciences and arts and humanities journals are underrepresented, while there is an overrepresentation of English-language journals. Besides the broader coverage of Scopus compared with WoS, it is also observed in the literature that citation counts tend to be higher in Scopus than in WoS (e.g., Haddow & Genoni, 2010; Kulkarni et al., 2009; Torres-Salinas et al., 2009).

Regarding the sensitivity of citation analyses to the choice between WoS and Scopus, results reported in the literature are somewhat mixed. Torres-Salinas et al. (2009) report that WoS and Scopus yield similar results for rankings of university departments. In an analysis in the field of information studies, Meho and Sugimoto (2009) observe that for smaller entities (e.g., journals, conference proceedings, and institutions) results based on WoS and Scopus are considerably different while for larger entities (e.g., research domains and countries) very similar results are obtained. This is in line with Archambault et al. (2009), who show that at the country level results based on WoS and Scopus are highly correlated.

**3.2. Comparing Google Scholar with Web of Science and Scopus**

There are a substantial number of studies in which Google Scholar is compared with WoS and sometimes also with Scopus. A number of studies report that Google Scholar outperforms WoS and Scopus in terms of coverage of publications. Meho and Yang (2007) analyze publications of library and information science researchers and report that, in comparison with WoS and Scopus, Google Scholar stands out in its coverage of conference proceedings and non-English language journals. Mingers and Lipitakis (2010) find that in the field of business and management Google Scholar has a much broader coverage than WoS, and they therefore conclude that WoS should not be used for measuring the impact of business and management research. Very similar observations are made by Amara and Landry (2012). Walters (2007) reports that Google Scholar has a substantially broader coverage than WoS in the field of later-life migration. Franceschet (2010a) compares WoS and Google Scholar in an analysis of computer science researchers and finds that Google Scholar identifies many more publications and citations than WoS. Similar results are obtained by García-Pérez (2010) in the field of psychology and by Wildgaard (2015) in the fields of astronomy, environmental science, philosophy, and public health. Harzing and Alakangas (2016) analyze researchers in five broad scientific disciplines and find that Google Scholar has a broader coverage than both WoS and Scopus. Kousha and Thelwall (2008) study the sources of citations that are counted by Google Scholar but not by WoS. They find that 70% of these citations originate from full-text scholarly sources available on the web.

On the other hand, there are studies indicating that the coverage of Google Scholar is not consistently broader than the coverage of WoS and Scopus. Mayr and Walter (2007) find that journals covered by WoS are not always covered by Google Scholar. The analysis of Bar-Ilan (2008b) suggests that Google Scholar has a broader coverage than WoS and Scopus in computer science and mathematics, but a worse coverage in high energy physics. Bornmann et al. (2009) report coverage problems of Google Scholar in the field of chemistry. They conclude that WoS and Scopus are more suitable than Google Scholar for research evaluation in chemistry. Bakkalbasi et al.

(2006) compare WoS, Scopus, and Google Scholar in the fields of oncology and condensed matter physics and indicate that none of the three databases consistently outperforms the others. A similar conclusion is reached by Kulkarni et al. (2009) based on an analysis of publications in general medical journals. Mikki (2010) presents a comparison of WoS and Google Scholar in the field of earth sciences and also reports that neither database has a consistently better performance than the other. It should be noted that a number of studies indicate substantial improvements in the coverage of Google Scholar over time (Chen, 2010; De Winter et al., 2014; Harzing, 2013b, 2014). This suggests that perhaps the results of earlier studies reporting coverage problems of Google Scholar may not be relevant anymore.

A different perspective on the comparison of Google Scholar with WoS and Scopus is offered by Li et al. (2010). These authors use WoS, Scopus, and Google Scholar to calculate citation impact indicators for a number of library and information science researchers, and they then correlate the indicators with judgments provided by experts. The three databases turn out to yield broadly similar results. Indicators calculated based on Scopus are most strongly correlated with expert judgment, while WoS-based indicators have the weakest correlation, but the differences are very small.

Various studies also investigate specific problems of Google Scholar. A general impression obtained from the literature is that Google Scholar suffers from a lack of quality control. Many inaccuracies in Google Scholar are reported in the literature. Jacsó (2005, 2006, 2010) for instance discusses problems related to content gaps, incorrect citation counts, and phantom data. The possibility of manipulating citation counts in Google Scholar is discussed by Beel and Gipp (2010), Labbé (2010), and López-Cózar et al. (2014). Google Scholar is also criticized for its lack of transparency (e.g., Jacsó, 2005; Wouters & Costas, 2012). It is unclear what is covered by Google Scholar and what is not. Researchers also point out that cleaning Google Scholar data can be very time consuming (Li et al., 2010; Meho & Yang, 2007).

**3.3. Social sciences and humanities**

Social sciences and humanities (SSH) research differs from research in the sciences in a number of fundamental ways. This is discussed in detail in literature reviews provided by Hicks (1999), Nederhof (2006), and Huang and Chang (2008).

Nederhof (2006) for instance lists the following key differences between SSH research and research in the sciences:

- SSH research has a stronger national and regional orientation (indicated for instance by publications in a national language rather than in English).
- SSH research is published less in journals and more in books.
- SSH research has a slower pace of theoretical development.
- SSH research is less collaborative.
- SSH research is directed more at a non-scholarly public.

As pointed out by Hicks (1999) and Nederhof (2006), because of the relatively strong national and regional orientation of SSH research, the coverage of SSH publications in WoS is limited. Many national and regional SSH journals are not covered by WoS. The significant role played by books in SSH research also contributes to the limited WoS coverage of SSH publications. Until recently, books were not covered at all by WoS.

The difficulties caused by the national and regional orientation of SSH research are emphasized by Archambault et al. (2006). They claim that WoS has a 20 to 25% overrepresentation of English language SSH journals. On the other hand, the difficulties caused by book publishing may diminish over time. Larivière et al. (2006) observe that journals play an increasingly important role in the social sciences (but not in the humanities). Also, WoS nowadays includes a Book Citation Index. This may make it possible to include books in citation analyses, although Gorraiz et al. (2013) conclude that the Book Citation Index at the moment should not yet be used for this purpose.

Further insight into the WoS coverage of SSH literature is provided by studies in which the complete SSH publication output of a country or region is compared with the output that is covered by WoS. Such studies are reported by Larivière and Macaluso (2011) for the province of Québec in Canada, by Engels et al. (2012) for the region of Flanders in Belgium, and by Sivertsen and Larsen (2012) for Norway. Larivière and Macaluso (2011) study the Érudit database, which is a database of journals from Québec, and report that in comparison with WoS this database includes about 30% more SSH publications from French-speaking universities in Québec. Based on an analysis of the VABB-SHW database, which is a database of publications authored by SSH researchers in Flanders, Engels et al. (2012) conclude that SSH researchers in Flanders increasingly publish their work in English, often in WoS covered journals, but they also report that there is no shift away from book publishing. The main observation made by Sivertsen and Larsen (2012), based on data on SSH publications from Norway, is that book publishing and domestic journal publishing show a concentration of many publications in a limited number of publication channels, which suggests that there are promising opportunities for obtaining a more comprehensive coverage of SSH literature. A comparison between the databases used in Flanders and Norway is presented by Ossenblok et al. (2012), who conclude that SSH researchers in Flanders display a stronger tendency to publish in WoS covered journals than Norwegian SSH researchers.

### 3.4. Conference proceedings

In certain fields, publications in conference proceedings play an important role. As discussed by Glänzel et al. (2006b), Lisée et al. (2008), and Vrettas and Sanderson (2015), this is especially the case in computer science and engineering. However, including conference proceedings publications in a citation analysis is difficult for a number of reasons. Below, two important difficulties are discussed.

The first difficulty is that little is known about the coverage of conference proceedings in WoS, Scopus, and Google Scholar. There is almost no work in which the three databases are compared. Exceptions are the studies by Meho and Rogers (2008) and Meho and Yang (2007), both of which have already been mentioned above. Meho and Rogers (2008) report that in the field of human-computer interaction Scopus has a broader coverage of conference proceedings than WoS. Meho and Yang (2007) find that in the field of library and information science Google Scholar outperforms WoS and Scopus in terms of its coverage of conference proceedings. Another study of the coverage of conference proceedings is reported by Michels and Fu (2014), but this study considers only the WoS database. Michels and Fu (2014) observe gaps in the coverage of important conferences in WoS.

The second difficulty in the use of conference proceedings publications in a citation analysis relates to the issue of double counting of work that is published both in a conference proceedings and in a journal. This issue is analyzed by Bar-Ilan (2010) and Michels and Fu (2014). As pointed out by Bar-Ilan (2010), double counting creates various problems. Most importantly, publication counts increase in an artificial way as a consequence of double counting, while citation counts per publication are likely to decrease.

## 4. Basic citation impact indicators

To organize the discussion on citation impact indicators, I start by distinguishing between a number of very basic indicators. These basic indicators are important because most indicators proposed in the literature can be seen as variants or extensions of these basic indicators.

As discussed in Section 3, the number of publications of a research unit (e.g., a researcher, a research group, a research institution, a country, or a journal) and the number of citations of these publications are likely to be different in different databases. However, from now on, it is simply assumed that one particular database is used and that citation impact indicators are calculated based on the publication and citation counts provided by this database.

The number of publications of a research unit also depends on the time period within which publications are counted. Likewise, the number of citations of a publication depends on the time period within which citations are counted. Instead of simply counting all publications and citations, one usually counts publications and citations only within a specific time period. The selection of the publications and citations that are included in the calculation of citation impact indicators is discussed in more detail in Section 5. For the moment, I simply assume that we work with a given set of publications and for each publication a given number of citations.

I distinguish between five basic citation impact indicators. These indicators are listed in Table 2. I will now briefly discuss each indicator:

- *Total number of citations*. The total number of citations of the publications of a research unit. As an example, consider a research unit with five publications, which have received 14, 12, 3, 1, and 0 citations. The total number of citations then equals 30.
- *Average number of citations per publication*. The average number of citations of the publications of a research unit. For the research unit in our example, the average number of citations per publication equals 30 / 5 = 6. Without doubt, the best-known indicator based on the idea of counting the average number of citations per publication is the journal impact factor, which counts the average number of citations received by the publications in a journal.

Indicators based on average citation counts are frequently used, but they are also criticized in the literature. Citation distributions tend to be highly skewed (e.g., Albarrán et al., 2011; Seglen, 1992), and therefore the average number of citations of a set of publications may be strongly influenced by one or a few highly cited publications. This is for instance observed by Aksnes and Sivertsen (2004) at the level of countries and by Waltman et al. (2012a) at the level of universities. Because of the skewness of citation distributions, it is sometimes suggested to replace or complement indicators based on average citation counts by alternative indicators (e.g., Aksnes & Sivertsen, 2004; Bornmann & Mutz, 2011; Leydesdorff & Opthof, 2011; Waltman et al., 2012a). Indicators based on the idea of counting highly cited publications are a frequently suggested alternative.

- *Number of highly cited publications*. The number of publications of a research unit that are considered to be highly cited, where a certain threshold needs to be chosen to determine whether a publication is counted as highly cited or not. For instance, using a threshold of ten citations, the research unit in our example has two highly cited publications. The idea of counting highly cited publications is suggested by for instance Martin and Irvine (1983) and Plomp (1990, 1994), and highly cited publications are sometimes seen as indications of scientific excellence (e.g., Bornmann, 2014; Tijssen et al., 2002). The i10-index reported by Google Scholar is based on the idea of counting highly cited publications.
- *Proportion of highly cited publications*. The proportion of the publications of a research unit that are considered to be highly cited. Using again a threshold of ten citations, the proportion of highly cited publications for the research unit in our example equals 2 / 5 = 0.4 (or 40%).
- *h-index*. The h-index (or Hirsch index) is defined as follows: A research unit has index h if h of its publications each have at least h citations and the other publications each have no more than h citations. For the research unit in our example, the h-index equals three. This is because the three most frequently cited publications each have at least three citations while the other two publications each have no more than three citations.

The h-index was introduced in 2005 (Hirsch, 2005) and has quickly become very popular. A large number of variants and extensions of the h-index have been proposed in the literature, of which the g-index (Egghe, 2006) is probably the one that is best known. Some counterintuitive properties of the h-index are highlighted by Waltman and Van Eck (2012a). In this review, no detailed discussion of the literature on the h-index and its variants is provided. Instead, the reader is referred to existing literature reviews (Alonso et al., 2009; Egghe, 2010; Norris & Oppenheim, 2010; Panaretos & Malesios, 2009).

Table 2. Five basic citation impact indicators, with a distinction between size-dependent and size-independent indicators.

| Size-dependent indicators | Size-independent indicators |
| --- | --- |
| Total number of citations | Average number of citations per publication |
| Number of highly cited publications | Proportion of highly cited publications |
| h-index | |

In Table 2, a distinction is made between size-dependent and size-independent indicators. Size-dependent indicators aim to provide an overall performance measure. When additional publications are obtained, these indicators will never decrease. On the other hand, size-independent indicators aim to provide an average performance measure per publication. These indicators may decrease when additional publications are obtained. Size-independent indicators are typically used to make comparisons between units that are of different size, for instance between a small and a large research group or between a small and a large university. Most citation impact indicators for journals, such as the impact factor, are also size independent. This is because when journals are compared, one often does not want the size of the journals (i.e., the number of publications in each journal) to have an effect on the comparison. It is usually more interesting to compare journals based on the average citation impact of their publications.

As can be seen in Table 2, the average number of citations per publication and the proportion of highly cited publications are size-independent indicators. These indicators have the total number of citations and the number of highly cited publications as their size-dependent counterparts. The h-index is also size dependent.

However, because of the special way in which publications and citations are combined in the h-index, this indicator does not have a size-independent counterpart.

It should be emphasized that this literature review is focused on citation impact indicators that are calculated exclusively based on 'output data', in particular data on publications and the citations received by these publications. Indicators that also take into account 'input data', for instance data on the number of researchers of a research unit or the amount of funding of a research unit, have received only limited attention in the literature, and therefore no specific discussion on these indicators is provided in this review. The advantage of using input data in addition to output data is that measurements can be obtained not only of citation impact but also of publication productivity. The use of input data is for instance advocated by Abramo and D'Angelo (2014), who provide a detailed discussion of the fractional scientific strength indicator. This indicator uses input and output data to provide combined measurements of publication productivity and citation impact.

## 5. Selection of publications and citations

In the calculation of citation impact indicators, a selection of the publications of a research unit is often made. Only the selected publications are taken into account in the calculation of the indicators. In many cases, only publications from a specific time period are considered, so a selection is made based on the year in which a publication appeared. A selection of publications can also be made in order to exclude certain types of publications, such as editorials, non-English language publications, or publications in national journals. In a similar way, a selection of the citations received by the publications of a research unit is sometimes made, and only the selected citations are considered in the calculation of citation impact indicators. This for instance allows self-citations to be left out. It also allows citations to be considered only within a specific time period after the appearance of a publication. This time period is usually referred to as the citation window. Below, a review is provided of the literature on excluding certain types of publications and citations from the calculation of citation impact indicators.

**5.1. Document type**

A common criterion for excluding publications from the calculation of citation impact indicators is based on the so-called document type of a publication. In WoS and Scopus, each publication has a document type. For instance, important document types in WoS are 'article', 'review', 'letter', 'editorial material', 'meeting abstract', and 'proceedings paper'. The main reason for excluding certain document types is that publications of different document types are hard to compare with each other. This problem is of limited significance in the case of basic size-dependent indicators such as the total number of citations or the h-index, but the problem is serious in the case of size-independent indicators such as the average number of citations per publication. For instance, consider a researcher who serves as editor of a journal and who now and then writes an editorial for his/her journal. Editorials are of a very different nature than ordinary research articles, and they therefore tend to be cited much less frequently. Using a size-independent indicator such as the average number of citations per publication, a researcher would essentially be penalized for writing editorials. This can be avoided by excluding editorials from the calculation of the average number of citations per publication.

In the literature, discussions on document types and their inclusion in or exclusion from the calculation of citation impact indicators mainly relate to the WoS database. González-Albo and Bordons (2011), Zhang and Glänzel (2012), and Harzing (2013a) discuss the 'proceedings paper' document type. Harzing (2013a) in addition also focuses on the 'review' document type. The document types 'letter' and 'editorial material' are discussed by, respectively, Van Leeuwen et al. (2007) and Van Leeuwen et al. (2013). For older literature on document types in the WoS database, I refer to Sigogneau (2000) and the references provided in this work.

**5.2. Language**

Another criterion for excluding publications from the calculation of citation impact indicators is the language in which a publication is written. Van Leeuwen et al. (2001) and Van Raan et al. (2011) suggest that in a comparative analysis of countries or research institutions publications not written in English should be excluded from the calculation of size-independent indicators. They show that non-English language publications on average receive fewer citations than English language publications, which they suggest is because many researchers cannot read publications that are not in English. Following this reasoning, they then argue that including non-English language publications creates a bias against countries in which researchers publish a lot in their own language.

**5.3. National vs. international journals**

Waltman and Van Eck (2013a, 2013b) go one step further and argue that not only non-English language publications should be excluded but all publications in journals that do not have a sufficiently strong international orientation. They present criteria for identifying these journals. The possibility of excluding non-international journals is also suggested by Moed (2002) and López-Illescas et al. (2009), based on the idea that international comparisons can best be made by considering only publications in the international scientific literature. Zitt et al. (2003) reason in a somewhat similar direction. They study the effect of excluding journals with a low citation impact, which are often journals with a national focus.

**5.4. Self-citations**

In addition to excluding certain types of publications from the calculation of citation impact indicators, it is also sometimes suggested to exclude certain types of citations, in particular self-citations. Self-citations can be defined at various levels, for instance at the journal level (i.e., a publication in a journal citing another publication in the same journal) or at the level of research institutions (i.e., a publication of an institution citing another publication of the same institution). However, in the literature, most attention is paid to self-citations at the level of authors. I therefore focus on these author self-citations.

Author self-citations are usually defined as citations for which the citing and the cited publication have at least one author in common (e.g., Aksnes, 2003; Glänzel et al., 2004). Although this is the most commonly used definition of author self-citations, some proposals for alternative definitions can be found in the literature. Costas et al. (2010) propose to distinguish between author self-citations and co-author self-citations (see also Schreiber, 2007, 2008a). From the point of view of a specific researcher, they define an author self-citation as a citation made by the researcher to his/her own work, while a co-author self-citation is defined as a citation made by a co-author of the researcher to one of their co-authored works. Another proposal is made by Schubert et al. (2006), who suggest a fractional author self-citation concept based on the degree of overlap between the set of authors of a citing publication and the set of authors of a cited publication. The problem of the algorithmic identification of self-citations is studied by Donner (2016). An important difficulty in the identification of self-citations is that the name of a researcher may be written in different ways in different publications.

Regardless of the definition of author self-citations that is adopted, one needs to decide whether author self-citations should be excluded from the calculation of citation impact indicators or not. At the macro level (e.g., countries), Aksnes (2003) and Glänzel and Thijs (2004) show that the effect of author self-citations is very small. Glänzel and Thijs (2004) therefore conclude that there is no need to exclude author self-citations. Aksnes (2003) argues that below the macro level author self-citations should preferably be excluded. At the meso level (e.g., research institutions), Thijs and Glänzel (2006) are in favor of presenting citation impact indicators both including and excluding author self-citations. As an alternative to excluding author self-citations, Glänzel et al. (2006a) suggest to offer supplementary indicators based on author self-citations. At the meso and micro level (e.g., individual researchers), Costas et al. (2010) consider non-self-citations to be the most relevant citations for evaluation purposes, but they emphasize that author self-citations also provide interesting information. At the micro level, Hirsch (2005) states that author self-citations should ideally be excluded, but he also claims that the h-index is not very sensitive to author self-citations, at least less sensitive than the total number of citations. Schreiber (2007) argues that Hirsch (2005) underestimates the sensitivity of the h-index to author self-citations. He prefers to exclude author self-citations from the calculation of the h-index, a position that is supported by Vinkler (2007) and Gianoli and Molina-Montenegro (2009). Schreiber (2008a) makes a similar point for the g-index, which he claims to be even more sensitive to author self-citations than the h-index. On the other hand, Engqvist and Frommen (2008, 2010), Henzinger et al. (2010), and Huang and Lin (2011) suggest that the sensitivity of the h-index to author self-citations is limited and, consequently, that there may be no need to exclude author self-citations.

Fowler and Aksnes (2007) suggest that excluding author self-citations from the calculation of citation impact indicators may not be sufficient, because author self-citations may serve as an advertisement of a researcher's work and may therefore have an increasing effect on the number of citations received from others. More precisely, they indicate that each author self-citation seems to yield an additional 3.65 citations from others. Their suggestion is that there might be a need for an explicit penalty on author self-citations. An earlier study by Medoff (2006), based on a more limited data set, does not find strong evidence of an 'advertisement effect' of author self-citations.

**5.5. Citation windows**

In the calculation of citation impact indicators, citations are sometimes taken into account only within a specific time period after the appearance of a publication, the so-called citation window. Adopting a certain citation window may cause both publications and citations to be excluded from the calculation of citation impact indicators. For instance, suppose we require publications to have a citation window of at least five years. For recent publications it is not possible to have a five-year citation window, and therefore this requirement implies that recent publications cannot be included in the calculation of citation impact indicators.

To illustrate the exclusion of citations from the calculation of citation impact indicators, suppose we want to compare the citation impact of publications from 2005 with the citation impact of publications from 2010. In order to make the comparison as fair as possible, we may want to have consistent citation windows. In the case of the 2010 publications, we may choose to count citations until the end of 2015. For consistency, in the case of the 2005 publications, we then need to count citations until the end of 2010. This means that for these publications citations received after 2010 are not included in the calculation of citation impact indicators.

At the level of individual publications, the choice of a citation window is studied by Adams (2005), Abramo et al. (2011), Waltman et al. (2011), and Wang (2013). Based on an analysis of publications in the life and physical sciences from the UK, Adams (2005) concludes that at the aggregate level citation statistics based on a short citation window (e.g., one year) correlate strongly with citation statistics based on a longer citation window (e.g., ten years). Abramo et al. (2011) analyze Italian publications in the sciences and claim that in all fields except for mathematics a citation window of two or three years seems sufficient to obtain robust citation impact indicators. Waltman et al. (2011) study the correlation between short-term and longer-term citation counts for biochemistry and molecular biology publications and for mathematics publications. They suggest that publications may need to have a citation window of at least one full year. Wang (2013) performs the most extensive study. For all publications from 1980 indexed in the WoS database, he analyzes the correlation between short-term and long-term citation counts. He argues that the choice of a citation window depends on a trade-off between accuracy and timeliness. Accuracy requires a longer citation window, while timeliness requires a shorter citation window. According to Wang (2013), there is no generally applicable rule for choosing citation windows.

The study of citation windows is closely related to the study of delayed recognition. Delayed recognition refers to the situation in which it takes a long time before the importance of a publication is recognized and the publication starts to receive significant numbers of citations. Based on an analysis of all WoS-indexed publications in the sciences from 1980, Glänzel et al. (2003) conclude that delayed recognition is an exceptional phenomenon that does not have much influence on citation impact indicators. A related analysis is presented by Van Raan (2004). He studies 'sleeping beauties in science', which are extreme examples of delayed recognition.

Levitt and Thelwall (2011) study citation windows from a different point of view. They argue that short citation windows have the problem that publications appearing in the first months of a year have a significant advantage over publications appearing in the last months. Levitt and Thelwall (2011) propose to address this problem by using a composite indicator in which the number of citations of a publication is combined with the impact factor of the journal in which the publication has appeared. This proposal is also discussed in Subsection 8.4.

Instead of studying the choice of a citation window at the level of individual publications, Costas et al. (2011) and Abramo et al. (2012a) analyze this choice at the level of researchers while Costas et al. (2013) and Abramo et al. (2012b) analyze it at the level of, respectively, research groups and universities. The different studies all reach a similar conclusion. Although there are some differences between fields (Abramo et al., 2012a, 2012b), the studies all find that citation impact indicators are relatively insensitive to the choice of a citation window and, consequently, that the use of short citation windows is justified.

A different conclusion is reached by Nederhof et al. (2012) in an analysis of publications in the life and physical sciences on the topic of space research. Nederhof et al. (2012) show that for these publications longer citation windows yield more favorable results than shorter citation windows. Based on this, they argue that space research publications require a citation window of at least five years.

## 6. Normalization

One of the key principles of citation analysis is that citation counts of publications from different fields should not be directly compared with each other. This is because there are large differences among fields in citation density, that is, in the average number of citations per publication. For instance, a biochemistry publication with 25 citations cannot be considered to have a higher citation impact than a mathematics publication with ten citations. There is a difference in citation density between biochemistry and mathematics of about an order of magnitude (Waltman et al., 2011b). Taking this into account, it needs to be concluded that the publication with the higher citation impact is actually the one in mathematics rather than the biochemistry one.

In addition to comparisons between publications from different fields, one should also be careful with comparisons between publications from different years. Even within the same field, a publication from 2005 with 25 citations cannot necessarily be considered to have a higher citation impact than a publication from 2010 with ten citations. Taking into account that the publication from 2005 has had five more years to attract citations, the conclusion may be that the publication with the higher citation impact is actually the one from 2010. This would for instance be a reasonable conclusion if we know that in the field of interest publications from 2005 on average have 40 citations while publications from 2010 on average have only five citations.

In a similar way, it is often argued that citation counts of publications of different document types, for instance the WoS document types 'article', 'letter', and 'review', should not be directly compared with each other, for instance because review articles tend to attract many more citations than ordinary research articles.

For practical purposes, there often is a need to make comparisons between publications that are from different fields or different years or that have different document types. Normalized citation impact indicators have been developed to make such comparisons. The idea of these indicators is to correct as much as possible for the effect of variables that one does not want to influence the outcomes of a citation analysis. In practice, one typically corrects for the effects of the field of a publication, the year in which a publication appeared, and sometimes also the document type of a publication. In principle, one could also correct for the effects of other variables, such as the number of authors of a publication or the length of a publication, but for these variables it is less clear whether performing a correction is desirable or not. Below, I review the literature on normalized citation impact indicators. My focus is on normalization for field differences. In general, normalization for differences in publication year and document type can be performed in a similar way.

For each of the five basic citation impact indicators presented in Table 2, it is possible to develop normalized variants. I start by discussing normalized variants of the average number of citations per publication. I then consider normalized variants of the proportion of highly cited publications. Normalized variants of the size-dependent counterparts of these two indicators can be obtained in a completely analogous way (e.g., Waltman et al., 2011a) and therefore I do not offer a further discussion on these indicators. One should be aware, however, that when size-dependent indicators have been normalized for differences among fields in citation density, these indicators are still sensitive to differences among fields in publication density. In the context of the h-index, the third size-dependent indicator listed in Table 2, there is some literature on the topic of normalization (Batista et al., 2006; Iglesias & Pecharromán, 2007; Kaur et al., 2013; Radicchi et al., 2008). However, since most work on normalization does not consider the h-index, I do not provide a further discussion of this literature.

**6.1. Normalized indicators based on average citation counts**

In the calculation of normalized variants of the average number of citations per publication, a key concept is the expected number of citations of a publication. The expected number of citations of a publication is defined as the average number of citations of all publications in the same field (and from the same year and of the same document type). When working with the WoS database, fields are often defined based on the WoS journal subject categories. WoS distinguishes between about 250 journal subject categories, most of which can be considered to represent a specific field of science, such as biochemistry, condensed matter physics, economics, mathematics, oncology, and sociology. Each journal covered by WoS belongs to one or more of these journal subject categories. Hence, based on the journal in which a publication has appeared, each publication indexed in WoS can be assigned to one or more journal subject categories, which then represent the fields to which the publication belongs.

Given the expected number of citations of a publication, the normalized citation score of the publication is calculated as the ratio of the actual number of citations of the publication and the expected number of citations. For a set of publications of a research unit, a normalized variant of the average number of citations per publication is obtained by taking the average of the normalized citation scores of the publications of the research unit. Table 3 provides a simple example. This example considers a research unit that has five publications. For each publication, both the actual and the expected number of citations is given (first two columns of Table 3). The normalized citation score of a publication is calculated by dividing the actual number of citations by the expected number of citations (last column of Table 3). Next, a normalized variant of the average number of citations per publication is obtained by averaging the normalized citation scores of the five publications. As shown in Table 3, the average normalized citation score equals 1.07. This score is somewhat above one, which indicates that on average the publications of the research unit have been cited above expectation.

Table 3. Example of the calculation of the average normalized citation score of a set of publications.

| Actual no. of cit. | Expected no. of cit. | Norm. cit. score |
|---|---|---|
| 14 | 21 | 0.67 |
| 12 | 4 | 3.00 |
| 3 | 2 | 1.50 |
| 1 | 5 | 0.20 |
| 0 | 2 | 0.00 |
| Average norm. cit. score: | | 1.07 |

Another normalized variant of the average number of citations per publication is obtained by first calculating, for a given set of publications, the total number of citations actually received and the expected total number of citations and by then taking the ratio of the actual and the expected total number of citations. For instance, in the case of the publications listed in Table 3, the actual total number of citations equals 30, while the expected total number of citations equals 34. Hence, the ratio of the actual and the expected total number of citations equals 30 / 34 = 0.88. The fact that the ratio is below one indicates that the total number of citations actually received is below expectation.

In the literature, there is no agreement which of the above two normalized variants of the average number of citations per publication is to be preferred. Most researchers nowadays seem to prefer the first variant, which is sometimes referred to as the average of ratios approach, over the second variant, which is sometimes called the ratio of averages approach. Using different arguments, Lundberg (2007), Opthof and Leydesdorff (2010), Van Raan et al. (2010), and Waltman et al. (2011a) claim that the average of ratios approach is more appropriate than the ratio of averages approach. However, Moed (2010b) and Vinkler (2012) present counterarguments in favor of the ratio of averages approach. Empirical comparisons between the two approaches are presented by Larivière and Gingras (2011), Waltman et al. (2011b), and Herranz and Ruiz-Castillo (2012). They conclude that the differences between the two approaches are small, especially at the level of countries and research institutions. In the context of the average of ratios approach, Smolinsky (2016) discusses different approaches to deal with overlapping fields, that is, publications belonging to more than one field.

In addition to the above discussion on averages of ratios versus ratios of averages, researchers have also studied various alternative approaches to calculate normalized citation scores. Lundberg (2007) suggests to apply a logarithmic transformation to citation counts and to normalize citation counts by calculating z-scores. Related ideas are also studied by Zhang et al. (2014), Fairclough and Thelwall (2015), and Thelwall (2016). Others have built on the work of Radicchi et al. (2008) and Radicchi and Castellano (2011), who start from the viewpoint that a proper normalization approach should result in normalized citation distributions that are universal across fields. Radicchi et al. (2008) conclude that normalization based on the ratio of the actual and the expected number of citations of a publication indeed yields the desired universality of citation distributions. However, Albarrán et al. (2011) and Waltman et al. (2012b) claim that this conclusion is too strong and that no perfect universality of citation distributions is obtained. Abramo et al. (2012c, 2012d) compare a number of normalization approaches and suggest that the best normalization is obtained by dividing the actual number of citations of a publication by the average number of citations of all publications that are in the same field and that have at least one citation. Radicchi and Castellano (2012b) introduce a normalization approach that is based on a transformation of citation counts by a two-parameter power-law function. Li et al. (2013) compare this normalization approach with a number of other approaches. Based on the degree to which the different approaches manage to create normalized citation distributions that are identical across fields, they conclude that the approach proposed by Radicchi and Castellano (2012b) has the best performance.

**6.2. Normalized indicators based on highly cited publications**

Normalized variants of the proportion of highly cited publications use a field-dependent threshold to determine whether a publication is counted as highly cited or not. The field-dependent threshold is usually chosen in such a way that the percentage of highly cited publications is the same in each field. This approach is proposed by Tijssen et al. (2002), who focus on the top 1% and the top 10% most highly cited publications in a field, and by Van Leeuwen et al. (2003), who consider the top 5% most highly cited publications. Nowadays, the idea of calculating the proportion of publications that belong to the top 10% most highly cited in their field plays an important role both in the CWTS Leiden Ranking and in the SCImago Institutions Rankings, which are the two most important bibliometric university rankings (Waltman et al., 2012a; Bornmann et al., 2012).

Choosing a citation threshold in such a way that a certain pre-specified percentage of the publications in a field, for instance 10% of the publications, are above the threshold is not entirely straightforward. It is usually not possible to obtain exactly the desired percentage of publications above the threshold. Depending on how the threshold is chosen, the percentage will be either somewhat too low or somewhat too high. The main cause of this difficulty is that there are often many publications in a field that all have the same number of citations. Because publications with the same number of citations will be either all below the threshold or all above the threshold, it becomes difficult to obtain exactly the desired percentage of publications above the threshold. There is some discussion in the literature on the best way to deal with this difficulty. Different approaches are proposed by, among others, Van Leeuwen et al. (2003), Pudovkin and Garfield (2009), Leydesdorff et al. (2011), Bornmann et al. (2012), and Waltman and Schreiber (2013). A summary of the different approaches is given by Waltman and Schreiber (2013), and an empirical comparison is presented by Schreiber (2013).

Leydesdorff et al. (2011) introduce a generalization of the idea of identifying a certain percentage of highly cited publications in each field. Instead of making a binary distinction between publications that are highly cited and publications that are not, Leydesdorff et al. (2011) suggest to define a number of classes of publications, where each class of publications is defined in terms of percentiles of the citation distribution of a field. For instance, the first class may include all publications whose number of citations is below the 50th percentile of the citation distribution of a field, the second class may include all publications whose number of citations is between the 50th and the 75th percentile, and so on. Leydesdorff et al. (2011) propose an indicator that values publications based on the class to which they belong, with publications in the lowest class having a value of one, publications in the second-lowest class having a value of two, etc. An approach that is somewhat similar to the approach of Leydesdorff et al. (2011) is presented by Glänzel (2013). Glänzel (2013) and Glänzel et al. (2014) also defines a number of classes of publications, but instead of percentiles he uses the method of characteristic scores and scales (Glänzel & Schubert, 1988) to define the classes. Publications belong to the lowest class if they have fewer citations than the average of their field, they belong to the second-lowest class if they do not belong to the lowest class and if they have fewer citations than the average of all publications that do not belong to the lowest class, and so on.

Another approach that focuses specifically on the highly cited publications in a field is proposed by Albarrán et al. (2011a, 2011b). They suggest a set of indicators that can be used to characterize the distribution of citations over the highly cited publications in a field. These indicators resemble indicators developed in the field of economics for characterizing income distributions.

**6.3. Choice of a field classification system**

Normalization of citation impact indicators, either of indicators based on average citation counts or of indicators based on highly cited publications, requires a classification system in which publications are assigned to fields. As explained above, the WoS journal subject categories are the most commonly used field classification system for normalization purposes. However, researchers have raised some important questions related to the choice of a classification system. These questions are for instance about the sensitivity of normalized indicators to the choice of a classification system and about the possibilities for using alternative classification systems instead of the WoS journal subject categories.

Zitt et al. (2005), Adams et al. (2008), Glänzel et al. (2009), and Colliander and Ahlgren (2011) study the sensitivity of normalized indicators to the aggregation level at which fields are defined. Zitt et al. (2005) and Adams et al. (2008) observe a lack of stability of normalized indicators with respect to the aggregation level at which normalization takes place. They argue that different aggregation levels provide different viewpoints and may all have a certain legitimacy. Glänzel et al. (2009) compare normalization at the level of WoS journal subject categories with normalization at higher aggregation levels defined according to the Leuven/Budapest field classification system (Glänzel and Schubert, 2003). Based on a macro level analysis of research institutions, they indicate that their preferred approach is to normalize at a relatively high aggregation level at which there are 60 fields. Colliander and Ahlgren (2011) perform an analysis of university departments and conclude that there are no substantial differences when instead of the WoS journal subject categories the 22 fields defined in the Essential Science Indicators are used for normalization purposes.

Other analyses of the suitability of the WoS journal subject categories for normalization purposes are reported by Van Eck et al. (2013) and Leydesdorff and Bornmann (2016). Van Eck et al. (2013) observe a strong heterogeneity in citation characteristics within medical subject categories, suggesting that the use of these subject categories for normalizing citation impact indicators may be problematic. Leydesdorff and Bornmann (2016) study the way in which two fields, namely library and information science and science and technology studies, are represented by WoS journal subject categories. They suggest that the WoS journal subject categories may be inappropriate for normalization purposes.

Researchers have proposed various improvements of and alternatives to the use of the WoS journal subject categories for normalizing citation impact indicators. Improvements are suggested by Glänzel et al. (1999) and Rons (2012). Glänzel et al. (1999) discuss the reassignment of publications in multidisciplinary journals (e.g., Nature and Science) to appropriate subject categories based on their references. Rons (2012) introduces the idea of exploiting the overlap of subject categories to obtain a more detailed classification system.

An obvious alternative to the use of the WoS journal subject categories is to replace them by an alternative field classification system. Proposals in this direction are made by Bornmann et al. (2008), Neuhaus and Daniel (2009), and Van Leeuwen and Calero-Medina (2012), who suggest the use of, respectively, Medical Subject Headings, Chemical Abstracts sections, and the EconLit classification system. An important limitation of these alternative classification systems is that each of them is restricted to a single field of science. Ruiz-Castillo and Waltman (2015) also propose the use of an alternative classification system, but instead of using an existing classification system they algorithmically construct their own classification system based on a large-scale analysis of citation relations between publications (Waltman & Van Eck, 2012). Their algorithmically constructed classification system covers all fields of science.

A critical perspective on the normalization of citation impact indicators is taken by Kostoff (2002) and Kostoff and Martinez (2005). They argue that the only meaningful normalization approach is to select for each publication a small number of thematically similar publications and to compare the number of citations of a publication with the number of citations received by the selected similar publications. According to Kostoff (2002) and Kostoff and Martinez (2005), selecting similar publications needs to be done manually by experts. Colliander (2015) proposes a somewhat similar approach, but instead of selecting similar publications manually he introduces an algorithm that selects similar publications based on shared references and shared terms. The idea of comparing publications with other similar publications selected based on shared references (i.e., bibliographic coupling) is also discussed by Schubert and Braun (1993, 1996). A somewhat similar idea at the level of journals instead of individual publications is proposed by Dorta-González et al. (2014).

**6.4. Alternative normalization approaches**

The normalization approaches discussed so far are based on the idea of comparing the number of citations of a publication with the number of citations of other publications that are considered to be in the same field. I now discuss some alternative normalization approaches that have been proposed in the literature. An attractive feature of these alternative normalization approaches is that they do not require a field classification system.

An important alternative normalization approach is given by the concept of citing-side normalization. Citing-side normalization is based on the idea that differences among fields in citation density are to a large extent caused by the fact that in some fields publications tend to have longer reference lists than in other fields. Citing-side normalization aims to normalize citation impact indicators by correcting for the effect of reference list length. The concept of citing-side normalization originates from Zitt and Small (2008). Different approaches to citing-side normalization are discussed by Zitt and Small (2008), Zitt (2010), Gómez-Sancho and Mancebón-Torrubia (2009), Moed (2010a), Leydesdorff and Opthof (2010), Leydesdorff and Bornmann (2011a),

Leydesdorff et al. (2013a), Waltman et al. (2013), and Glänzel et al. (2011). Empirical comparisons between citing-side normalization and traditional cited-side normalization are presented by Glänzel et al. (2011), Radicchi and Castellano (2012a), Leydesdorff et al. (2013b), and Waltman and Van Eck (2013a, 2013b). Radicchi and Castellano (2012a) and Leydesdorff et al. (2013b) conclude that cited-side normalization performs better than citing-side normalization, but Sirtes (2012) criticizes the methodology on which this conclusion is based. Waltman and Van Eck (2013a, 2013b) reach the opposite conclusion and suggest that citing-side normalization may outperform cited-side normalization. However, their conclusion is challenged by Ruiz-Castillo (2014).

Ideas similar to citing-side normalization are also suggested by Nicolaisen and Frandsen (2008), Kosmulski (2011), Franceschini et al. (2012), and Franceschini and Maisano (2014). However, these authors propose to perform a normalization based on the reference list length of cited publications, while citing-side normalization is based on the reference list length of citing publications.

Recursive citation impact indicators offer another alternative normalization approach. These indicators give different weights to citations depending on their source. The higher the citation impact of the source of a citation, the higher the weight of the citation. Like indicators based on citing-side normalization, recursive citation impact indicators correct for the effect of reference list length. The idea of recursive citation impact indicators originates from Pinski and Narin (1976). The introduction of the well-known PageRank algorithm (Brin & Page, 1998) has led to a renewed interest in recursive citation impact indicators. Overviews of the literature on these indicators are provided by Waltman and Yan (2014) and Fragkiadaki and Evangelidis (2014).

## 7. Counting methods

Science is becoming increasingly collaborative. Various studies have for instance shown a continuously increasing trend in the average number of authors per publication (e.g., Gazni et al., 2012; Larivière et al., 2015; Persson et al., 2004; Wuchty et al., 2007). Extreme examples of large-scale scientific collaboration can be found in high energy physics and in certain biomedical fields, where publications sometimes include several hundreds of authors (e.g., Cronin, 2001).

With increasing numbers of authors per publication, it becomes more and more difficult to properly allocate the credits of a publication to the individual authors. Citation impact indicators often allocate the full credits of a publication to each individual author. This approach is known as full counting, whole counting, integer counting, or total counting. For instance, if a publication with five authors has been cited ten times, each author is considered to have ten citations. Hence, overall 50 citations are allocated to the five authors. It is clear that this approach has an inflationary effect, since citations received by publications with multiple authors are counted multiple times. This is sometimes considered undesirable, and therefore various alternative approaches to dealing with multi-author publications have been proposed in the literature. Below, I first discuss the fractional counting method. I then review a number of other counting methods suggested in the literature.

**7.1. Fractional counting**

In the fractional counting method, the credits of a publication are fractionally allocated to the authors of the publication. Each author receives an equal share of the credits. For instance, in the case of a publication with five authors and ten citations, each author receives one fifth of the credits of the publication, which means that each author is allocated two citations.

When working at the level of countries or institutions rather than individual researchers, there are different ways in which fractional counting can be implemented. For instance, in the case of a publication co-authored by three US researchers and one UK researcher, one possibility is to allocate the publication with weight 0.75 to the US and with weight 0.25 to the UK. Another possibility is to allocate the publication to each country with a weight of 0.5. Detailed discussions of the different possibilities are provided by Gauffriau et al. (2007) and Waltman and Van Eck (2015). These researchers also present proposals of a systematic terminology that can be used to distinguish between different counting methods.

Comparisons between full and fractional counting in analyses at the level of countries are reported by, among others, Rinia et al. (1993), Gauffriau and Larsen (2005), Moed (2005), Gauffriau et al. (2008), Huang et al. (2011), Aksnes et al. (2012), and Waltman and Van Eck (2015). Gauffriau et al. (2008) also provide references to earlier work in which full and fractional counting are compared. Empirical comparisons between the two counting methods show that fractional counting yields lower citation scores than full counting. This is because publications co-authored by multiple countries on average receive more citations than publications authored by a single country. In the fractional counting method, publications co-authored by multiple countries have less weight, and therefore fractional counting yields lower citation scores than full counting. There is no general consensus on which of the two counting methods is to be preferred. It can be argued that full and fractional counting measure different concepts (participation vs. contribution) and both provide useful information. This perspective is emphasized by Moed (2005). However, in most studies in which full and fractional counting are compared, a preference for fractional counting is indicated (Aksnes et al., 2012; Gauffriau & Larsen, 2005; Huang et al., 2011; Rinia et al., 1993; Waltman & Van Eck, 2015). Full counting is often criticized because it provides non-additive statistics, with for instance the sum of the number of publications of each country in the world being larger than the total number of publications worldwide.

At the institutional level, full and fractional counting are compared by Waltman et al. (2012a), Lin et al. (2013), and Waltman and Van Eck (2015). These researchers express a preference for fractional counting over full counting. Waltman et al. (2012a) and Waltman and Van Eck (2015) argue that full counting may lead to invalid comparisons across fields, even when working with normalized indicators. From a somewhat different perspective, this problem is also studied by Perianes-Rodríguez and Ruiz-Castillo (2015).

At the level of individual researchers, Price (1981) argues that fractional counting is preferable over full counting. Lindsey (1980) presents an overview of bibliometric analyses at the level of individual researchers reported in the sociology of science literature. Most studies turn out to use full counting, but Lindsey (1980) argues that fractional counting is preferable. The introduction of the h-index has led to a renewed interested in counting methods at the level of individual researchers. Fractional counting variants of the h-index are studied by Egghe (2008) and Schreiber (2008b, 2008c, 2009a). The same researchers also investigate fractional counting variants of the g-index (Egghe, 2008; Schreiber, 2009b, 2010a, 2010b).

**7.2. Other counting methods**

A common objection against fractional counting is that distributing the credits of a publication equally over all authors may not be fair. Some authors may have contributed more than others, and ideally this should be reflected in the way in which credit is allocated to authors. In the literature, various approaches have been proposed for allocating credit to the authors of a publication based on their position in the author list. This is based on the idea that the position of an author in the author list of a publication provides an indication of the contribution made by the author, with the first author typically being regarded as the most important contributor. Of course, this idea is not valid in fields in which the authors of a publication tend to be ordered alphabetically. This phenomenon of alphabetical authorship is studied by Frandsen and Nicolaisen (2010) and Waltman (2012). Waltman (2012) finds that alphabetical authorship is common in mathematics, economics, and high energy physics. A review of the literature on authorship order is provided by Marušić et al. (2011).

A simple approach to allocate credit to authors based on their position in the author list of a publication is to give the full credits of a publication to the first author and to give no credits at all to the other authors. This approach is known as first-author counting or straight counting. First-author counting has been studied in country-level analyses (Gauffriau et al., 2008; Huang et al., 2011; Rinia et al., 1993; Schubert et al., 1989; Waltman & Van Eck, 2015), institutional-level analyses (Lin et al., 2013), and analyses at the level of individual researchers (Lange, 2001; Lindsey, 1980). Instead of allocating the credits of a publication to the first author, researchers have also investigated the idea of allocating the credits to the corresponding author (Huang et al., 2011; Lin et al., 2013; Moya-Anegón et al., 2013; Waltman & Van Eck, 2015). Another possibility is to allocate credit both to the first and to corresponding author of a publication. Hu et al. (2010) explore this possibility in the context of the h-index. It should be noted that the concepts of first author and corresponding author can be somewhat ambiguous. Hu (2009) draws attention to the fact that an increasing number of publications have multiple first authors ('equal first authorship') or multiple corresponding authors.

Various more complex approaches to allocate credit to authors based on their position in the author list of a publication have been proposed. These approaches assign weights to the authors of a publication. The weight of an author depends on the position of the author in the author list and on the total number of authors of the publication. The typical idea is to assign the highest weight to the first author, followed by the second author, the third author, and so on. The total weight of all authors of a publication usually equals one. The weight of an author can then be interpreted as the share of the credits of the publication that are allocated to that author. Weights can be assigned to authors in many different ways, and therefore a number of different weighted counting methods have been introduced in the literature. These include harmonic counting (Hagen, 2008, 2010, 2013, 2014a, 2014b, 2015; Hodge & Greenberg, 1981; Jian & Xiaoli, 2013), arithmetic counting (Abbas, 2011; Egghe et al., 2000; Van Hooydonk, 1997), also known as proportional counting, geometric counting (Egghe et al., 2000), the counting method of Assimakis and Adam (2010) based on the golden number, and the axiomatic counting method of Stallings et al. (2013). Table 4 illustrates the differences between these approaches by showing the weights assigned to the authors of a publication with five authors. Other weighted counting methods are proposed by Lukovits and Vinkler (1995), Trueba and Guerrero (2004), Liu and Fang (2012a, 2012b), Abramo et al. (2013), and Kim and Diesner (2014). Comparisons of different methods are presented by Kim and Kim (2015) and Xu et al. (in press).

Table 4. Weights assigned to the authors of a publication with five authors. The weights are determined based on harmonic counting, arithmetic counting, geometric counting, and the counting methods of Assimakis and Adam (2010) and Stallings et al. (2013).

|  | 1st author | 2nd author | 3rd author | 4th author | 5th author |
|---|---|---|---|---|---|
| Harmonic counting | 0.438 | 0.219 | 0.146 | 0.109 | 0.088 |
| Arithmetic counting | 0.333 | 0.267 | 0.200 | 0.133 | 0.067 |
| Geometric counting | 0.516 | 0.258 | 0.129 | 0.065 | 0.032 |
| Assimakis and Adam (2010) | 0.618 | 0.236 | 0.090 | 0.034 | 0.021 |
| Stallings et al. (2013) | 0.457 | 0.257 | 0.157 | 0.090 | 0.040 |

A critical perspective on weighted counting methods is presented by Kosmulski (2012). He argues that weighted counting methods fail to take into consideration the situation of group leaders, who in many cases are listed as the last author of a publication. When weights are assigned to authors based on their position in the author list of a publication, group leaders often will not be assigned a correct weight. A possible solution to this problem is proposed by Aziz and Rozing (2013), who introduce a counting method that gives most weight to the first and the last author of a publication and least weight to the author in the middle of the author list. Other researchers have suggested weighted counting methods that do not depend at all on the order of the authors of a publication. These approaches therefore do not suffer from the group leader problem discussed by Kosmulski (2012). One approach is suggested by Tol (2011), who proposes to assign weights to the authors of a publication based on each author's past performance. Another approach, suggested by Shen and Barabasi (2014), assigns weights to the authors of a publication by taking into account co-citation relations between the publication and each author's earlier work.

Discussions on weighted counting methods often take place in the context of the h-index (Abbas, 2011; Galam, 2011; Hagen, 2008; Jian & Xiaoli, 2013; Liu & Fang, 2012a, 2012b). However, in addition to weighted counting methods, researchers have also proposed alternative ways of correcting the h-index for the effect of co-authorship. Batista et al. (2006) and Wan et al. (2007) suggest to divide the h-index by a correction factor that depends on the number of co-authors someone has. A more complex proposal is made by Hirsch (2010), who introduces a variant of the h-index referred to as the h-bar-index. A publication contributes to someone's h-bar-index only if it also contributes to the h-bar-index of each of the co-authors of the publication.

## 8. Citation impact indicators for journals

The discussion in the previous sections has focused on citation impact indicators in general. In this section, I focus specifically on citation impact indicators for journals. A separate section is devoted to this topic because of the large amount of attention it receives in the literature. I refer to Glänzel and Moed (2002), Rousseau (2002), Bar-Ilan (2008a), and Haustein (2012) for earlier overviews of the literature on indicators of the citation impact of journals. Empirical comparisons of various citation impact indicators for journals are reported by Bollen et al. (2009), Leydesdorff (2009), and Elkins et al. (2010).

**8.1. Basic citation impact indicators for journals**

The best-known indicator of the citation impact of journals is the impact factor (Garfield, 1972). The impact factor of a journal equals the ratio of on the one hand the number of citations given in a particular year to publications in the journal in the previous two years and on the other hand the number of publications in the journal in the previous two years. For instance, if a journal published a total of 100 publications in 2012 and 2013 and if these publications were cited 200 times in 2014, the impact factor of the journal equals 200 / 100 = 2. Hence, the impact factor essentially equals the average number of citations of the publications of a journal. However, the interpretation of the impact factor as a journal's average number of citations per publication is not entirely correct. This is because in the numerator of the impact factor citations to publications of all document types are counted while in the denominator only publications of specific document types (i.e., so-called citable documents) are included (Moed & Van Leeuwen, 1995, 1996).

There is a large amount of literature on the impact factor. Here I mention only a few selected works. Garfield (1996, 2006) discusses the history, interpretation, and proper use of the impact factor from the perspective of its inventor. More details on the history of the impact factor are provided by Bensman (2007) and Archambault and Larivière (2009). The impact factor causes a lot of debate. Some of the discussion on the impact factor is summarized by Bar-Ilan (2008a). Recently, discussion took place in a special issue of Scientometrics (Braun, 2012). This discussion was triggered by a critical paper about the impact factor by Vanclay (2012). It should be noted, however, that part of the debate about the impact factor is not so much about the indicator itself but more about the way in which the indicator is used for research assessment purposes. In particular, there is much criticism on the use of the impact factor for assessing individual publications (and their authors) based on the journal in which they have appeared. I will get back to this below.

In addition to the classical impact factor based on citations to publications in the previous two years, there is also a five-year impact factor, which takes into account citations to publications in the previous five years. The five-year impact factor addresses the criticism that in some fields the two-year citation window of the classical impact factor is too short (e.g., Glänzel & Schoepflin, 1995; Moed et al., 1998). An empirical comparison between the two-year and the five-year impact factor is presented by Campanario (2011). The two-year and the five-year impact factor are both available in the Journal Citation Reports produced by Thomson Reuters. The Journal Citation Reports also include the immediacy index, an indicator of the frequency at which the publications in a journal are cited in the year in which they appeared. Some other citation impact indicators included in the Journal Citation Reports will be discussed below.

Various other basic citation impact indicators for journals have been proposed in the literature, either as an alternative or as a complement to the impact factor. Ingwersen et al. (2001), Frandsen and Rousseau (2005), and Ingwersen (2012) discuss a so-called diachronic variant of the impact factor. In the ordinary synchronic impact factor, citations in a single year to publications in multiple earlier years are counted. In the diachronic impact factor, citations in multiple years to publications in a single year are counted, for instance citations in 2012, 2013, and 2014 to publications in 2012. Another variant of the impact factor is introduced by Sombatsompop et al. (2004) and Rousseau (2005). They propose an ordinary synchronic impact factor, but instead of considering publications in a fixed two-year time period their proposed impact factor considers publications in a flexible journal-dependent time period. The longer it takes for the publications in a journal to be cited, the longer the time period in which publications are taken into consideration in the impact factor of the journal. In this way, the impact factor is adjusted to the specific citation characteristics of a journal. A somewhat similar proposal is presented by Dorta-González and Dorta-González (2013). Thelwall and Fairclough (2015) introduce yet another variant of the impact factor. In order to obtain a more stable indicator, they propose to calculate the impact factor using a geometric average instead of an arithmetic one. Other basic citation impact indicators for journals suggested in the literature include the share of (un)cited publications (Markpin et al., 2008; Van Leeuwen & Moed, 2005), the median number of citations (Calver & Bradley, 2009), and the h-index (Braun et al., 2006; Harzing & Van der Wal, 2009). It should be noted that unlike most indicators for journals the h-index is size dependent. Journals with more publications tend to have higher h-indices.

**8.2. Normalized citation impact indicators for journals**

The citation impact indicators for journals discussed above do not correct for differences in citation density among fields. To address this limitation, a large number of normalized citation impact indicators have been proposed in the literature.

The simplest proposal is made by Pudovkin and Garfield (2004). They suggest a normalized citation impact indicator for journals that is based on the rank of a journal within its WoS subject category when journals are ordered by their impact factor. For instance, if a journal has the 10th highest impact factor within a subject category that includes 200 journals, the journal is assigned a score of (approximately) 0.95, indicating that 95% of the journals in the subject category have a lower impact factor.

Building on their earlier work (Moed et al., 1998, 1999), Van Leeuwen and Moed (2002) propose a citation impact indicator for journals that is normalized for field, publication year, and document type. Normalization is implemented by comparing the actual number of citations of each publication in a journal with the expected number of citations, where the expected number of citations of a publication is given by the average number of citations of all publications in the same field and publication year and of the same document type. Related proposals on normalized citation impact indicators for journals are presented by Sen (1992), Marshakova-Shaikevich (1996), Sombatsompop and Markpin (2005), and Vieira and Gomes (2011). These proposals all use the WoS subject categories to define fields. Mutz and Daniel (2012a, 2012b) also suggest an approach for normalizing citation impact indicators for journals. Their focus is mainly on normalization for document type rather than normalization for field. Leydesdorff and Bornmann (2011b) and Wagner and Leydesdorff (2012) introduce a normalized citation impact indicator for journals that, unlike most indicators for journals, is not based on average citation counts per publication. Following the ideas developed by Leydesdorff et al. (2011), the proposed indicator values the publications in a journal based on their position within the citation distribution of the field. Glänzel (2011) proposes a somewhat similar idea based on the method of characteristic scores and scales (Glänzel & Schubert, 1988).

During recent years, another approach to the normalization of citation impact indicators for journals has been developed. This is the citing-side normalization approach introduced by Zitt and Small (2008). The SNIP (source normalized impact per paper) indicator provided in Scopus is based on citing-side normalization. The original version of this indicator is presented by Moed (2010a). The version that is currently included in Scopus is described by Waltman et al. (2013). A comparison between the two versions of the SNIP indicator is presented by Moed (2016), and a critical perspective on the current version of the SNIP indicator is provided by Mingers (2014). I refer to Subsection 6.4 for a further discussion of the literature on citing-side normalization. Most of this literature focuses on indicators for journals.

**8.3. Recursive citation impact indicators for journals**

As already mentioned in Subsection 6.4, recursive citation impact indicators give different weights to citations depending on their source, with citations originating from a high-impact source having more weight than citations originating from a low-impact source. The idea for instance is that being cited in *Nature* or *Science* should be valued more than being cited in an obscure journal that almost no one knows about. The first proposal of a recursive citation impact indicator for journals is made by Pinski and Narin (1976). A more recent proposal, inspired by the well-known PageRank algorithm (Brin & Page, 1998), is made by Bollen et al. (2006). Recursive citation impact indicators for journals are included both in the Journal Citation Reports and in Scopus. The Journal Citation Reports include the eigenfactor and article influence indicators (Bergstrom, 2007; West et al., 2010a), while Scopus includes the SCImago journal rank (SJR) indicator (González-Pereira et al., 2010; Guerrero-Bote & Moya-Anegón, 2012). I now discuss these indicators in more detail. A more extensive overview of the literature on recursive citation impact indicators for journals is provided by Waltman and Yan (2014).

Like the impact factor, the article influence indicator is obtained by calculating the average number of citations of the publications in a journal. However, unlike the impact factor, the article influence indicator gives more weight to citations from high-impact journals than to citations from low-impact journals. The size-dependent counterpart of the article influence indicator is referred to as the eigenfactor indicator. This indicator is proportional to the product of the number of publications of a journal and the article influence indicator. Hence, the eigenfactor indicator takes the size of a journal into account and therefore favors larger journals over smaller ones. The article influence indicator and the eigenfactor indicator have the special property that self-citations at the level of journals are not counted. Citations given by a journal to itself are ignored in the calculation of the indicators. For a further discussion on the article influence indicator and the eigenfactor indicator, including empirical comparisons with other citation impact indicators for journals, I refer to Davis (2008), West et al. (2010b), Franceschet (2010b, 2010c, 2010d), and Walters (2014).

The SJR indicator has two versions, the original version introduced by González-Pereira et al. (2010) and the revised version discussed by Guerrero-Bote and Moya-Anegón (2012). The revised version is the one that is currently included in Scopus.

The SJR indicator is fairly similar to the article influence indicator, although its mathematical definition is more complex. A special feature of the revised SJR indicator is that the weight of a citation depends not only on the citation impact of the citing journal but also on a measure of the thematic closeness of the citing and the cited journal. A citation from a citing journal that is thematically close to the cited journal is given more weight than a citation from a more distant citing journal.

**8.4. Citation impact of journals vs. citation impact of individual publications**

In research assessments, there often is a tendency to evaluate publications based on the citation impact of the journal in which they have appeared. Especially the impact factor is often used for this purpose. Evaluating publications based on the impact factor of the journal in which they have appeared is attractive because impact factors are easily available, more easily than statistics on the number of times individual publications have been cited. Impact factors therefore often serve as a substitute for publication-level citation statistics.

Many bibliometricians reject the use of the impact factor and other journal-level indicators for evaluating individual publications. The most important argument against this practice is that the citation impact of a journal offers only a weak predictor of the citation impact of individual publications in the journal. This is because the distribution of citations over the publications in a journal tends to be highly skewed, with for instance 20% of the publications receiving 60% of the citations. The average number of citations of the publications in a journal is therefore determined mainly by a small proportion of highly cited publications, and most publications in a journal have a citation impact that is substantially below the citation impact of the journal as a whole. Hence, the citation impact of a journal is not representative of the citation impact of a typical publication in the journal. This argument against the use of journal-level indicators for evaluating individual publications has received widespread support in the literature. Especially the work by Seglen (1992, 1994, 1997) on this topic has been influential. The inventor of the impact factor also warns against the use of this indicator for evaluating individual publications (Garfield, 1996, 2006). Recently, the San Francisco Declaration on Research Assessment (http://am.ascb.org/dora/), which strongly argues against the use of the impact factor in the assessment of individual publications and their authors, received a lot of support in the scientific community.

Some researchers argue that indicators of the citation impact of journals may be useful in evaluating very recent publications. In the case of very recent publications, the number of citations received provides hardly any information, simply because there has been almost no opportunity for these publications to be cited. The citation impact of the journal in which a publication has appeared may then be seen as an interesting alternative source of information. This line of reasoning is followed by Abramo et al. (2010) and Levitt and Thelwall (2011). Abramo et al. (2010) argue that in certain fields very recent publications can better be evaluated based on the impact factor of their journal than based on their individual number of citations. Levitt and Thelwall (2011) suggest to evaluate recent publications using a composite indicator that takes into account both the impact factor of the journal in which a publication has appeared and the number of citations received by the publication. In line with this suggestion, Stern (2014) reports that in the prediction of the long-term number of citations of recent publications the impact factor offers useful complementary information to the short-term number of citations. On the other hand, Lozano et al. (2012) claim that since 1990 the relation between the impact factor and the number of citations of individual publications has been weakening, suggesting that the use of the impact factor as a substitute for publication-level citation statistics is becoming more and more problematic.

## 9. Recommendations for future research

In this paper, I have provided a review of the literature on citation impact indicators, focusing on bibliographic databases based on which indicators can be calculated, the selection of publications and citations to be included in the calculation of indicators, the normalization of indicators, the different counting methods that can be used to handle co-authored publications, and the topic of citation impact indicators for journals.

I have attempted to review the literature on citation impact indicators in an objective way, without putting special emphasis on my personal ideas on these indicators. However, I will conclude this review by providing a more personal perspective on research on citation impact indicators. I will do so by offering four recommendations for future research.

*Recommendation 1: Do not introduce new citation impact indicators unless they have a clear added value relative to existing indicators.*

During the past decade, a very large number of new citation impact indicators have been introduced in the literature. Some of these indicators were discussed in this review, but many of them were not discussed because they have been introduced in the h-index literature that was not covered in detail in this review. Given the large number of citation impact indicators that have already been proposed, there is little need for having even more indicators. Many indicators proposed in the literature have different definitions but nevertheless provide essentially the same information. This is not surprising, since the information that can be obtained from citation counts is inherently limited, and therefore it is unlikely that there is a practical need to have more than a handful of indicators. New citation impact indicators should be proposed only if convincing arguments can be presented of their added value relative to existing indicators.

*Recommendation 2: Pay more attention to the theoretical foundation of citation impact indicators.*

Many citation impact indicators lack a strong theoretical foundation. The consequences of specific choices in the construction an indicator are not always fully understood, and the assumptions on which an indicator is based are not always stated in a clear and explicit way. For instance, many field-normalized citation impact indicators have been introduced in the literature, but often these indicators lack a mathematical framework that makes clear how the idea of field normalization is interpreted and under which assumptions field normalization is indeed achieved. Instead, field-normalized indicators are usually tested only in an empirical way. Likewise, many new citation impact indicators have been proposed, especially in the h-index literature, that have only a superficial justification and that lack a deeper reflection on the concept they intend to measure. In research on citation impact indicators, more attention should be paid to the theoretical foundation of indicators.

*Recommendation 3: Pay more attention to the way in which citation impact indicators are being used in practice.*

The literature on citation impact indicators has a strong technical focus. For instance, during the past years, a number of quite complex technical ideas on the problem of normalization for field differences have been introduced in the literature. Having technically sophisticated citation impact indicators is absolutely essential in some situations. At the same time, however, it is important to think about the practical use of indicators and to make sure that the technical criteria based on which indicators are constructed match well with the needs and expectations of the end users of indicators. This is something which has received less attention in the literature. There are a number of important questions that require more research. For instance, how are citation impact indicators being used in practice? How are these indicators interpreted and what types of conclusions are drawn from them? How is a concept such as field normalization understood by end users of citation impact indicators? Do end users overestimate the accuracy of field-normalized indicators? Answering questions such as these will be important in the development of best practices for the use of citation impact indicators (cf. Hicks et al., 2015). It will require technically focused research on citation impact indicators to be complemented by research that takes a more qualitative approach.

*Recommendation 4: Exploit new data sources to obtain more sophisticated measurements of citation impact.*

There are important ongoing developments in scientific publishing that are likely to create opportunities to obtain more advanced measurements of citation impact. One development is the introduction of more sophisticated ways in which the contributions that authors have made to a publication can be specified, for instance by having group authors in addition to ordinary authors, by distinguishing between authors, contributors, and guarantors, or by providing author contribution statements. These improved ways of specifying author contributions may offer new possibilities to address the credit allocation problem discussed in Section 7. Another major development is the increase in open access publishing, and related to this, the increase in the availability of the full text of scientific publications (instead of only the bibliographic meta data). The availability of full text data enables the construction of more advanced citation impact indicators, for instance indicators that take into account the number of times a publication is referenced in a citing publication (e.g., Ding et al., 2013; Wan & Liu, 2014; Zhu et al., 2015), the location (e.g., introduction, methods, results, or discussion) where a publication is referenced in a citing publication (e.g., Bertin et al., 2016; Ding et al., 2013; Hu et al., 2013), or even the context in which a publication is referenced (i.e., the sentences in a citing publication around the reference to a cited publication). Bibliometricians and scientometricians should broaden their perspective on citation analysis in order to take advantage of the opportunities offered by new data sources.

## Acknowledgments

I would like to thank Martijn Visser for his help in reviewing the literature on bibliographic databases. I am grateful to Giovanni Abramo, Lutz Bornmann, Wouter Gerritsma, Rüdiger Mutz, Antonio Perianes-Rodriguez, Javier Ruiz-Castillo, and two referees for their comments on an earlier version of this paper. This research was partly funded by the Higher Education Funding Council for England (HEFCE).

## References

Abbas, A. M. (2011). Weighted indices for evaluating the quality of research with multiple authorship. *Scientometrics*, *88*(1), 107-131.

Abramo, G., Cicero, T., & D'Angelo, C. A. (2011). Assessing the varying level of impact measurement accuracy as a function of the citation window length. *Journal of Informetrics*, *5*(4), 659-667.

Abramo, G., Cicero, T., & D'Angelo, C. A. (2012a). A sensitivity analysis of researchers' productivity rankings to the time of citation observation. *Journal of Informetrics*, *6*(2), 192-201.

Abramo, G., Cicero, T., & D'Angelo, C. A. (2012b). A sensitivity analysis of research institutions' productivity rankings to the time of citation observation. *Journal of Informetrics*, *6*(2), 298-306.

Abramo, G., Cicero, T., & D'Angelo, C. A. (2012c). How important is choice of the scaling factor in standardizing citations? *Journal of Informetrics*, *6*(4), 645-654.

Abramo, G., Cicero, T., & D'Angelo, C. A. (2012d). Revisiting the scaling of citations for research assessment. *Journal of Informetrics*, *6*(4), 470-479.

Abramo, G., & D'Angelo, C. A. (2014). How do you define and measure research productivity? *Scientometrics*, *101*(2), 1129-1144.

Abramo, G., D'Angelo, C. A., & Di Costa, F. (2010). Citations versus journal impact factor as proxy of quality: Could the latter ever be preferable? *Scientometrics*, *84*(3), 821-833.

Abramo, G., D'Angelo, C. A., & Rosati, F. (2013). The importance of accounting for the number of co-authors and their order when assessing research performance at the individual level in the life sciences. *Journal of Informetrics*, *7*(1), 198-208.

Adams, J. (2005). Early citation counts correlate with accumulated impact. *Scientometrics*, *63*(3), 567-581.

Adams, J., Gurney, K., & Jackson, L. (2008). Calibrating the zoom—A test of Zitt's hypothesis. *Scientometrics*, *75*(1), 81-95.

Aksnes, D. W. (2003). A macro study of self-citation. *Scientometrics*, *56*(2), 235-246.

Aksnes, D. W., Schneider, J. W., & Gunnarsson, M. (2012). Ranking national research systems by citation indicators. A comparative analysis using whole and fractionalised counting methods. *Journal of Informetrics*, *6*(1), 36-43.

Aksnes, D. W., & Sivertsen, G. (2004). The effect of highly cited papers on national citation indicators. *Scientometrics*, *59*(2), 213-224.

Albarrán, P., Crespo, J. A., Ortuño, I., & Ruiz-Castillo, J. (2011). The skewness of science in 219 sub-fields and a number of aggregates. *Scientometrics*, *88*(2), 385-397.

Albarrán, P., Ortuño, I., & Ruiz-Castillo, J. (2011a). The measurement of low- and high-impact in citation distributions: Technical results. *Journal of Informetrics*, *5*(1), 48-63.

Albarrán, P., Ortuño, I., & Ruiz-Castillo, J. (2011b). High- and low-impact citation measures: Empirical applications. *Journal of Informetrics*, *5*(1), 122-145.Alonso, S., Cabrerizo, F. J., Herrera-Viedma, E., & Herrera, F. (2009). h-Index: A review focused in its variants, computation and standardization for different scientific fields. *Journal of Informetrics*, *3*(4), 273-289.

Amara, N., & Landry, R. (2012). Counting citations in the field of business and management: Why use Google Scholar rather than the Web of Science. *Scientometrics*, *93*(3), 553-581.

Archambault, É., Campbell, D., Gingras, Y., & Larivière, V. (2009). Comparing bibliometric statistics obtained from the Web of Science and Scopus. *Journal of the American Society for Information Science and Technology*, *60*(7), 1320-1326.

Archambault, É., & Larivière, V. (2009). History of the journal impact factor: Contingencies and consequences. *Scientometrics*, *79*(3), 635-649.

Archambault, É., Vignola-Gagné, É., Côté, G., Larivière, V., & Gingras, Y. (2006). Benchmarking scientific output in the social sciences and humanities: The limits of existing databases. *Scientometrics*, *68*(3), 329-342.

Assimakis, N., & Adam, M. (2010). A new author's productivity index: p-index. *Scientometrics*, *85*(2), 415-427.

Aziz, N. A., & Rozing, M. P. (2013). Profit (p)-index: The degree to which authors profit from co-authors. *PLoS ONE*, *8*(4), e59814.

Bakkalbasi, N., Bauer, K., Glover, J., & Wang, L. (2006). Three options for citation tracking: Google Scholar, Scopus and Web of Science. *Biomedical Digital Libraries*, *3*(1), 7.

Bar-Ilan, J. (2008a). Informetrics at the beginning of the 21st century—A review. *Journal of Informetrics*, *2*(1), 1-52.

Bar-Ilan, J. (2008b). Which h-index? A comparison of WoS, Scopus and Google Scholar. *Scientometrics*, *74*(2), 257-271.

Bar-Ilan, J. (2010). Web of Science with the Conference Proceedings Citation Indexes: The case of computer science. *Scientometrics*, *83*(3), 809-824.

Bartol, T., Budimir, G., Dekleva-Smrekar, D., Pusnik, M., & Juznic, P. (2014). Assessment of research fields in Scopus and Web of Science in the view of national research evaluation in Slovenia. *Scientometrics*, *98*(2), 1491-1504.

Batista, P. D., Campiteli, M. G., & Kinouchi, O. (2006). Is it possible to compare researchers with different scientific interests? *Scientometrics*, *68*(1), 179-189.

Beel, J., & Gipp, B. (2010). Academic search engine spam and Google Scholar's resilience against it. *Journal of Electronic Publishing*, *13*(3).

Bensman, S. J. (2007). Garfield and the impact factor. *Annual Review of Information Science and Technology*, *41*(1), 93-155.

Bergstrom, C. T. (2007). Eigenfactor: Measuring the value and prestige of scholarly journals. *College and Research Libraries News*, *68*(5), 314–316.

Bertin, M., Atanassova, I., Gingras, Y., & Larivière, V. (2016). The invariant distribution of references in scientific articles. *Journal of the Association for Information Science and Technology*, *67*(1), 164-177.

Bollen, J., Rodriquez, M. A., & Van de Sompel, H. (2006). Journal status. *Scientometrics*, *69*(3), 669-687.

Bollen, J., Van de Sompel, H., Hagberg, A., & Chute, R. (2009). A principal component analysis of 39 scientific impact measures. *PLOS ONE*, *4*(6), e6022.

Bornmann, L. (2014). How are excellent (highly cited) papers defined in bibliometrics? A quantitative analysis of the literature. *Research Evaluation*, *23*(2), 166-173.

Bornmann, L., & Daniel, H.-D. (2008). What do citation counts measure? A review of studies on citing behavior. *Journal of Documentation*, *64*(1), 45-80.

Bornmann, L., De Moya Anegón, F., & Leydesdorff, L. (2012). The new excellence Indicator in the World Report of the SCImago Institutions Rankings 2011. *Journal of Informetrics*, *6*(2), 333-335.

Bornmann, L., Marx, W., Schier, H., Rahm, E., Thor, A., & Daniel, H. D. (2009). Convergent validity of bibliometric Google Scholar data in the field of chemistry—Citation counts for papers that were accepted by Angewandte Chemie International Edition or rejected but published elsewhere, using Google Scholar, Science Citation Index, Scopus, and Chemical Abstracts. *Journal of informetrics*, *3*(1), 27-35.

Bornmann, L., & Mutz, R. (2011). Further steps towards an ideal method of measuring citation performance: The avoidance of citation (ratio) averages in field-normalization. *Journal of Informetrics*, *5*(1), 228-230.

Bornmann, L., Mutz, R., Neuhaus, C., & Daniel, H. D. (2008). Citation counts for research evaluation: Standards of good practice for analyzing bibliometric data and presenting and interpreting results. *Ethics in Science and Environmental Politics*, *8*(1), 93-102.

Braun, T. (2012). Editorial. *Scientometrics*, *92*(2), 207-208.

Braun, T., Glänzel, W., & Schubert, A. (2006). A Hirsch-type index for journals. *Scientometrics*, *69*(1), 169-173.

Brin, S., & Page, L. (1998). The anatomy of a large-scale hypertextual web search engine. *Computer Networks and ISDN Systems*, *30*(1), 107-117.

Calver, M. C., & Bradley, J. S. (2009). Should we use the mean citations per paper to summarise a journal's impact or to rank journals in the same field? *Scientometrics*, *81*(3), 611-615.

Campanario, J. M. (2011). Empirical study of journal impact factors obtained using the classical two-year citation window versus a five-year citation window. *Scientometrics*, *87*(1), 189-204.

Cavacini, A. (2015). What is the best database for computer science journal articles? *Scientometrics*, *102*(3), 2059-2071.

Chen, X. (2010). Google Scholar's dramatic coverage improvement five years after debut. *Serials Review*, *36*(4), 221-226.

Colliander, C. (2015). A novel approach to citation normalization: A similarity-based method for creating reference sets. *Journal of the Association for Information Science and Technology*, *66*(3), 489-500.

Colliander, C., & Ahlgren, P. (2011). The effects and their stability of field normalization baseline on relative performance with respect to citation impact: A case study of 20 natural science departments. *Journal of Informetrics*, *5*(1), 101-113.

Costas, R., Van Leeuwen, T. N., & Bordons, M. (2010). Self-citations at the meso and individual levels: Effects of different calculation methods. *Scientometrics*, *82*(3), 517-537.

Costas, R., Van Leeuwen, T. N., & Van Raan, A. F. J. (2011). The "Mendel syndrome" in science: Durability of scientific literature and its effects on bibliometric analysis of individual scientists. *Scientometrics*, *89*(1), 177-205.

Costas, R., Van Leeuwen, T. N., & Van Raan, A. F. J. (2013). Effects of the durability of scientific literature at the group level: Case study of chemistry research groups in the Netherlands. *Research Policy*, *42*(4), 886-894.

Cronin, B. (2001). Hyperauthorship: A postmodern perversion or evidence of a structural shift in scholarly communication practices? *Journal of the American Society for Information Science and Technology*, *52*(7), 558-569.

Davis, P. M. (2008). Eigenfactor: Does the principle of repeated improvement result in better estimates than raw citation counts? *Journal of the American Society for Information Science and Technology*, *59*(13), 2186-2188.

De Bellis, N. (2009). *Bibliometrics and citation analysis: From the Science Citation Index to cybermetrics*. Scarecrow Press.

De Rijcke, S., Wouters, P., Rushforth, A., Franssen, T., & Hammarfelt, B. (2015). *Evaluation practices and effects of indicator use – A literature review*. Manuscript submitted for publication.

De Solla Price, D. (1981). Multiple authorship. *Science*, *212*, 986.

De Winter, J. C., Zadpoor, A. A., & Dodou, D. (2014). The expansion of Google Scholar versus Web of Science: A longitudinal study. *Scientometrics*, *98*(2), 1547-1565.

Delgado López-Cózar, E., Robinson-García, N., & Torres-Salinas, D. (2014). The Google Scholar experiment: How to index false papers and manipulate bibliometric indicators. *Journal of the Association for Information Science and Technology*, *65*(3), 446-454.

Ding, Y., Liu, X., Guo, C., & Cronin, B. (2013). The distribution of references across texts: Some implications for citation analysis. *Journal of Informetrics*, *7*(3), 583-592.

Donner, P. (2016). Enhanced self-citation detection by fuzzy author name matching and complementary error estimates. *Journal of the Association for Information Science and Technology*, *67*(3), 662-670.

Dorta-González, P., & Dorta-González, M. I. (2013). Impact maturity times and citation time windows: The 2-year maximum journal impact factor. *Journal of Informetrics*, *7*(3), 593-602.

Dorta-González, P., Dorta-González, M. I., Santos-Peñate, D. R., & Suárez-Vega, R. (2014). Journal topic citation potential and between-field comparisons: The topic normalized impact factor. *Journal of Informetrics*, *8*(2), 406-418.

Egghe, L. (2006). Theory and practise of the g-index. *Scientometrics*, *69*(1), 131-152.

Egghe, L. (2008). Mathematical theory of the h- and g-index in case of fractional counting of authorship. *Journal of the American Society for Information Science and Technology*, *59*(10), 1608-1616.

Egghe, L. (2010). The Hirsch index and related impact measures. *Annual Review of Information Science and Technology*, *44*(1), 65-114.

Egghe, L., Rousseau, R., & Van Hooydonk, G. (2000). Methods for accrediting publications to authors or countries: Consequences for evaluation studies. *Journal of the American Society for Information Science*, *51*(2), 145-157.

Elkins, M. R., Maher, C. G., Herbert, R. D., Moseley, A. M., & Sherrington, C. (2010). Correlation between the journal impact factor and three other journal citation indices. *Scientometrics*, *85*(1), 81-93.

Engels, T. C., Ossenblok, T. L., & Spruyt, E. H. (2012). Changing publication patterns in the social sciences and humanities, 2000–2009. *Scientometrics*, *93*(2), 373-390.

Engqvist, L., & Frommen, J. G. (2008). The h-index and self-citations. *Trends in Ecology and Evolution*, *23*(5), 250-252.

Engqvist, L., & Frommen, J. G. (2010). New insights into the relationship between the h-index and self-citations? *Journal of the American Society for Information Science and Technology*, *61*(7), 1514-1516.

Fairclough, R., & Thelwall, M. (2015). More precise methods for national research citation impact comparisons. *Journal of Informetrics*, *9*(4), 895-906.

Fowler, J. H., & Aksnes, D. W. (2007). Does self-citation pay? *Scientometrics*, *72*(3), 427-437.

Fragkiadaki, E., & Evangelidis, G. (2014). Review of the indirect citations paradigm: Theory and practice of the assessment of papers, authors and journals. *Scientometrics*, *99*(2), 261-288.

Franceschet, M. (2010a). A comparison of bibliometric indicators for computer science scholars and journals on Web of Science and Google Scholar. *Scientometrics*, *83*(1), 243-258.

Franceschet, M. (2010b). The difference between popularity and prestige in the sciences and in the social sciences: A bibliometric analysis. *Journal of Informetrics*, *4*(1), 55-63.

Franceschet, M. (2010c). Journal influence factors. *Journal of Informetrics*, *4*(3), 239-248.

Franceschet, M. (2010d). Ten good reasons to use the eigenfactor metrics. *Information Processing and Management*, *46*(5), 555-558.

Franceschini, F., Galetto, M., Maisano, D., & Mastrogiacomo, L. (2012). The success-index: An alternative approach to the h-index for evaluating an individual's research output. *Scientometrics*, *92*(3), 621-641.

Franceschini, F., & Maisano, D. (2014). Sub-field normalization of the IEEE scientific journals based on their connection with technical societies. *Journal of Informetrics*, *8*(3), 508-533.

Franceschini, F., Maisano, D., & Mastrogiacomo, L. (2013). A novel approach for estimating the omitted-citation rate of bibliometric databases with an application to the field of bibliometrics. *Journal of the American Society for Information Science and Technology*, *64*(10), 2149-2156.

Franceschini, F., Maisano, D., & Mastrogiacomo, L. (2014). Scientific journal publishers and omitted citations in bibliometric databases: Any relationship? *Journal of Informetrics*, *8*(3), 751-765.

Franceschini, F., Maisano, D., & Mastrogiacomo, L. (2015a). Errors in DOI indexing by bibliometric databases. *Scientometrics*, *102*(3), 2181-2186.

Franceschini, F., Maisano, D., & Mastrogiacomo, L. (2015b). Influence of omitted citations on the bibliometric statistics of the major manufacturing journals. *Scientometrics*, *103*(3), 1083-1122.

Franceschini, F., Maisano, D., & Mastrogiacomo, L. (2016). The museum of errors/horrors in Scopus. *Journal of Informetrics*, *10*(1), 174-182.

Franceschini, F., Maisano, D., & Mastrogiacomo, L. (in press). Do Scopus and WoS correct "old" omitted citations? *Scientometrics*.

Frandsen, T. F., & Nicolaisen, J. (2010). What is in a name? Credit assignment practices in different disciplines. *Journal of Informetrics*, *4*(4), 608-617.

Frandsen, T. F., & Rousseau, R. (2005). Article impact calculated over arbitrary periods. *Journal of the American Society for Information Science and Technology*, *56*(1), 58-62.

Galam, S. (2011). Tailor based allocations for multiple authorship: A fractional gh-index. *Scientometrics*, *89*(1), 365-379.

García-Pérez, M. A. (2010). Accuracy and completeness of publication and citation records in the Web of Science, PsycINFO, and Google Scholar: A case study for the computation of h indices in psychology. *Journal of the American Society for Information Science and Technology*, *61*(10), 2070-2085.

García-Pérez, M. A. (2011). Strange attractors in the Web of Science database. *Journal of Informetrics*, *5*(1), 214-218.

Garfield, E. (1972). Citation analysis as a tool in journal evaluation. *Science*, *178*, 471-479.

Garfield, E. (1996). How can impact factors be improved? *British Medical Journal*, *313*, 411.

Garfield, E. (2006). The history and meaning of the journal impact factor. *JAMA*, *295*(1), 90-93.

Gauffriau, M., & Larsen, P. O. (2005). Counting methods are decisive for rankings based on publication and citation studies. *Scientometrics*, *64*(1), 85-93.

Gauffriau, M., Larsen, P. O., Maye, I., Roulin-Perriard, A., & Von Ins, M. (2007). Publication, cooperation and productivity measures in scientific research. *Scientometrics*, *73*(2), 175-214.

Gauffriau, M., Larsen, P. O., Maye, I., Roulin-Perriard, A., & Von Ins, M. (2008). Comparisons of results of publication counting using different methods. *Scientometrics*, *77*(1), 147-176.

Gavel, Y., & Iselid, L. (2008). Web of Science and Scopus: A journal title overlap study. *Online Information Review*, *32*(1), 8-21.

Gazni, A., Sugimoto, C. R., & Didegah, F. (2012). Mapping world scientific collaboration: Authors, institutions, and countries. *Journal of the American Society for Information Science and Technology*, *63*(2), 323-335.

Gianoli, E., & Molina-Montenegro, M. A. (2009). Insights into the relationship between the h-index and self-citations. *Journal of the American Society for Information Science and Technology*, *60*(6), 1283-1285.

Glänzel, W. (2011). The application of characteristic scores and scales to the evaluation and ranking of scientific journals. *Journal of Information Science*,*37*(1), 40-48.

Glänzel, W. (2013). High-end performance or outlier? Evaluating the tail of scientometric distributions. *Scientometrics*, *97*(1), 13-23.

Glänzel, W., Debackere, K., Thijs, B., & Schubert, A. (2006a). A concise review on the role of author self-citations in information science, bibliometrics and science policy. *Scientometrics*, *67*(2), 263-277.

Glänzel, W., & Moed, H. F. (2002). Journal impact measures in bibliometric research. *Scientometrics*, *53*(2), 171-193.

Glänzel, W., Schlemmer, B., Schubert, A., & Thijs, B. (2006b). Proceedings literature as additional data source for bibliometric analysis. *Scientometrics*, *68*(3), 457-473.

Glänzel, W., Schlemmer, B., & Thijs, B. (2003). Better late than never? On the chance to become highly cited only beyond the standard bibliometric time horizon. *Scientometrics*, *58*(3), 571-586.

Glänzel, W., & Schoepflin, U. (1995). A bibliometric study on ageing and reception processes of scientific literature. *Journal of Information Science*, *21*(1), 37-53.

Glänzel, W., & Schubert, A. (1988). Characteristic scores and scales in assessing citation impact. *Journal of Information Science*, *14*(2), 123-127.

Glänzel, W., & Schubert, A. (2003). A new classification scheme of science fields and subfields designed for scientometric evaluation purposes. *Scientometrics*, *56*(3), 357-367.

Glänzel, W., Schubert, A., & Czerwon, H. J. (1999). An item-by-item subject classification of papers published in multidisciplinary and general journals using reference analysis. *Scientometrics*, *44*(3), 427-439.

Glänzel, W., Schubert, A., Thijs, B., & Debackere, K. (2011). A priori vs. a posteriori normalisation of citation indicators. The case of journal ranking. *Scientometrics*, *87*(2), 415-424.

Glänzel, W., & Thijs, B. (2004). The influence of author self-citations on bibliometric macro indicators. *Scientometrics*, *59*(3), 281-310.

Glänzel, W., Thijs, B., & Debackere, K. (2014). The application of citation-based performance classes to the disciplinary and multidisciplinary assessment in national comparison and institutional research assessment. *Scientometrics*, *101*(2), 939-952.

Glänzel, W., Thijs, B., & Schlemmer, B. (2004). A bibliometric approach to the role of author self-citations in scientific communication. *Scientometrics*, *59*(1), 63-77.

Glänzel, W., Thijs, B., Schubert, A., & Debackere, K. (2009). Subfield-specific normalized relative indicators and a new generation of relational charts: Methodological foundations illustrated on the assessment of institutional research performance. *Scientometrics*, *78*(1), 165-188.

Gómez-Sancho, J. M., & Mancebón-Torrubia, M. J. (2009). The evaluation of scientific production: Towards a neutral impact factor. *Scientometrics*, *81*(2), 435-458.

González-Albo, B., & Bordons, M. (2011). Articles vs. proceedings papers: Do they differ in research relevance and impact? A case study in the library and information science field. *Journal of Informetrics*, *5*(3), 369-381.

González-Pereira, B., Guerrero-Bote, V. P., & Moya-Anegón, F. (2010). A new approach to the metric of journals' scientific prestige: The SJR indicator. *Journal of informetrics*, *4*(3), 379-391.

Gorraiz, J., Melero-Fuentes, D., Gumpenberger, C., & Valderrama-Zurián, J.-C. (2016). Availability of digital object identifiers (DOIs) in Web of Science and Scopus. *Journal of Informetrics*, *10*(1), 98-109.

Gorraiz, J., Purnell, P. J., & Glänzel, W. (2013). Opportunities for and limitations of the Book Citation Index. *Journal of the American Society for Information Science and Technology*, *64*(7), 1388-1398.

Guerrero-Bote, V. P., & Moya-Anegón, F. (2012). A further step forward in measuring journals' scientific prestige: The SJR2 indicator. *Journal of Informetrics*, *6*(4), 674-688.

Haddow, G., & Genoni, P. (2010). Citation analysis and peer ranking of Australian social science journals. *Scientometrics*, *85*(2), 471-487.

Hagen, N. T. (2008). Harmonic allocation of authorship credit: Source-level correction of bibliometric bias assures accurate publication and citation analysis. *PLOS ONE*, *3*(12), e4021.

Hagen, N. T. (2010). Harmonic publication and citation counting: sharing authorship credit equitably–not equally, geometrically or arithmetically. *Scientometrics*, *84*(3), 785-793.

Hagen, N. T. (2013). Harmonic coauthor credit: A parsimonious quantification of the byline hierarchy. *Journal of Informetrics*, *7*(4), 784-791.

Hagen, N. T. (2014a). Counting and comparing publication output with and without equalizing and inflationary bias. *Journal of Informetrics*, *8*(2), 310-317.

Hagen, N. T. (2014b). Reversing the byline hierarchy: The effect of equalizing bias on the accreditation of primary, secondary and senior authors. *Journal of Informetrics*, *8*(3), 618-627.

Hagen, N. T. (2015). Contributory inequality alters assessment of academic output gap between comparable countries. *Journal of Informetrics*, *9*(3), 629-641.

Harzing, A. W. (2010). *The publish or perish book*. Tarma Software Research.

Harzing, A. W. (2013a). Document categories in the ISI Web of Knowledge: Misunderstanding the social sciences? *Scientometrics*, *94*(1), 23-34.

Harzing, A. W. (2013b). A preliminary test of Google Scholar as a source for citation data: A longitudinal study of Nobel prize winners. *Scientometrics*, *94*(3), 1057-1075.

Harzing, A. W. (2014). A longitudinal study of Google Scholar coverage between 2012 and 2013. *Scientometrics*, *98*(1), 565-575.

Harzing, A. W., & Alakangas, S. (2016). Google Scholar, Scopus and the Web of Science: A longitudinal and cross-disciplinary comparison. *Scientometrics*, *106*(2), 787-804.

Harzing, A. W., & Van der Wal, R. (2009). A Google Scholar h-index for journals: An alternative metric to measure journal impact in economics and business. *Journal of the American Society for Information Science and Technology*, *60*(1), 41-46.

Haustein, S. (2012). *Multidimensional journal evaluation*. De Gruyter.

Henzinger, M., Suñol, J., & Weber, I. (2010). The stability of the h-index. *Scientometrics*, *84*(2), 465-479.

Herranz, N., & Ruiz-Castillo, J. (2012). Sub-field normalization in the multiplicative case: Average-based citation indicators. *Journal of Informetrics*, *6*(4), 543-556.

Hicks, D. (1999). The difficulty of achieving full coverage of international social science literature and the bibliometric consequences. *Scientometrics*, *44*(2), 193-215.

Hicks, D., Wouters, P., Waltman, L., De Rijcke, S., & Rafols, I. (2015). The Leiden Manifesto for research metrics. *Nature*, *520*, 429-431.

Hirsch, J. E. (2005). An index to quantify an individual's scientific research output. *Proceedings of the National Academy of Sciences of the United States of America*, *102*(46), 16569-16572.

Hirsch, J. E. (2010). An index to quantify an individual's scientific research output that takes into account the effect of multiple coauthorship. *Scientometrics*, *85*(3), 741-754.

Hodge, S. E., & Greenberg, D. A. (1981). Publication credit. *Science*, *213*, 950.

Hu, X. (2009). Loads of special authorship functions: Linear growth in the percentage of "equal first authors" and corresponding authors. *Journal of the American Society for Information Science and Technology*, *60*(11), 2378-2381.

Hu, X., Rousseau, R., & Chen, J. (2010). In those fields where multiple authorship is the rule, the h-index should be supplemented by role-based h-indices. *Journal of Information Science*, *36*(1), 73-85.

Hu, Z., Chen, C., & Liu, Z. (2013). Where are citations located in the body of scientific articles? A study of the distributions of citation locations. *Journal of Informetrics*, *7*(4), 887-896.

Huang, M. H., & Chang, Y. W. (2008). Characteristics of research output in social sciences and humanities: From a research evaluation perspective. *Journal of the American Society for Information Science and Technology*, *59*(11), 1819-1828.

Huang, M. H., & Lin, W. Y. C. (2011). Probing the effect of author self-citations on h index: A case study of environmental engineering. *Journal of Information Science*, *37*(5), 453-461.

Huang, M. H., Lin, C. S., & Chen, D. Z. (2011). Counting methods, country rank changes, and counting inflation in the assessment of national research productivity and impact. *Journal of the American Society for Information Science and Technology*, *62*(12), 2427-2436.

Iglesias, J. E., & Pecharromán, C. (2007). Scaling the h-index for different scientific ISI fields. *Scientometrics*, *73*(3), 303-320.

Ingwersen, P. (2012). The pragmatics of a diachronic journal impact factor. *Scientometrics*, *92*(2), 319-324.

Ingwersen, P., Larsen, B., Rousseau, R., & Russell, J. (2001). The publication-citation matrix and its derived quantities. *Chinese Science Bulletin*, *46*(6), 524-528.

Jacsó, P. (2005). Google Scholar: The pros and the cons. *Online Information Review*, *29*(2), 208-214.

Jacsó, P. (2006). Deflated, inflated and phantom citation counts. *Online Information Review*, *30*(3), 297-309.

Jacsó, P. (2010). Metadata mega mess in Google Scholar. *Online Information Review*, *34*(1), 175-191.

Jian, D., & Xiaoli, T. (2013). Perceptions of author order versus contribution among researchers with different professional ranks and the potential of harmonic counts for encouraging ethical co-authorship practices. *Scientometrics*, *96*(1), 277-295.

Kaur, J., Radicchi, F., & Menczer, F. (2013). Universality of scholarly impact metrics. *Journal of Informetrics*, *7*(4), 924-932.

Kawashima, H., & Tomizawa, H. (2015). Accuracy evaluation of Scopus Author ID based on the largest funding database in Japan. *Scientometrics*, *103*(3), 1061-1071.

Khabsa, M., & Giles, C. L. (2014). The number of scholarly documents on the public web. *PLOS ONE*, *9*(5), e93949.

Kim, J., & Diesner, J. (2014). A network-based approach to coauthorship credit allocation. *Scientometrics*, *101*(1), 587-602.

Kim, J., & Kim, J. (2015). Rethinking the comparison of coauthorship credit allocation schemes. *Journal of Informetrics*, *9*(3), 667-673.

Kosmulski, M. (2011). Successful papers: A new idea in evaluation of scientific output. *Journal of Informetrics*, *5*(3), 481-485.

Kosmulski, M. (2012). The order in the lists of authors in multi-author papers revisited. *Journal of Informetrics*, *6*(4), 639-644.

Kostoff, R. N. (2002). Citation analysis of research performer quality. *Scientometrics*, *53*(1), 49-71.

Kostoff, R. N., & Martinez, W. L. (2005). Is citation normalization realistic? *Journal of information science*, *31*(1), 57-61.

Kousha, K., & Thelwall, M. (2008). Sources of Google Scholar citations outside the Science Citation Index: A comparison between four science disciplines. *Scientometrics*, *74*(2), 273-294.

Kousha, K. & Thelwall, M. (2015). Web indicators for research evaluation, part 3: Books and non-standard outputs. *El Profesional de la Información*, *24*(6), 724-736.

Kulkarni, A. V., Aziz, B., Shams, I., & Busse, J. W. (2009). Comparisons of citations in Web of Science, Scopus, and Google Scholar for articles published in general medical journals. *JAMA*, *302*(10), 1092-1096.

Labbé, C. (2010). Ike Antkare, one of the great stars in the scientific firmament. *ISSI Newsletter*, *6*(2), 48-52.

Lange, L. L. (2001). Citation counts of multi-authored papers—First-named authors and further authors. *Scientometrics*, *52*(3), 457-470.

Larivière, V., Archambault, É., Gingras, Y., & Vignola-Gagné, É. (2006). The place of serials in referencing practices: Comparing natural sciences and engineering with social sciences and humanities. *Journal of the American Society for Information Science and Technology*, *57*(8), 997-1004.

Larivière, V., & Gingras, Y. (2011). Averages of ratios vs. ratios of averages: An empirical analysis of four levels of aggregation. *Journal of Informetrics*, *5*(3), 392-399.

Larivière, V., Gingras, Y., Sugimoto, C. R., & Tsou, A. (2015). Team size matters: Collaboration and scientific impact since 1900. *Journal of the Association for Information Science and Technology*, *66*(7), 1323-1332.

Larivière, V., & Macaluso, B. (2011). Improving the coverage of social science and humanities researchers' output: The case of the Érudit journal platform. *Journal of the American Society for Information Science and Technology*, *62*(12), 2437-2442.

Larsen, P. O., & Von Ins, M. (2010). The rate of growth in scientific publication and the decline in coverage provided by Science Citation Index. *Scientometrics*, *84*(3), 575-603.

Levitt, J. M., & Thelwall, M. (2011). A combined bibliometric indicator to predict article impact. *Information Processing and Management*, *47*(2), 300-308.

Leydesdorff, L. (2009). How are new citation-based journal indicators adding to the bibliometric toolbox? *Journal of the American Society for Information Science and Technology*, *60*(7), 1327-1336.

Leydesdorff, L., & Bornmann, L. (2011a). How fractional counting of citations affects the impact factor: Normalization in terms of differences in citation potentials among fields of science. *Journal of the American Society for Information Science and Technology*, *62*(2), 217-229.

Leydesdorff, L., & Bornmann, L. (2011b). Integrated impact indicators compared with impact factors: An alternative research design with policy implications. *Journal of the American Society for Information Science and Technology*, *62*(11), 2133-2146.

Leydesdorff, L., & Bornmann, L. (2016). The operationalization of "fields" as WoS subject categories (WCs) in evaluative bibliometrics: The cases of "library and information science" and "science & technology studies". *Journal of the Association for Information Science and Technology*, *67*(3), 707-714.

Leydesdorff, L., Bornmann, L., Mutz, R., & Opthof, T. (2011). Turning the tables on citation analysis one more time: Principles for comparing sets of documents. *Journal of the American Society for Information Science and Technology*, *62*(7), 1370-1381.

Leydesdorff, L., & Opthof, T. (2010). Scopus's source normalized impact per paper (SNIP) versus a journal impact factor based on fractional counting of citations. *Journal of the American Society for Information Science and Technology*, *61*(11), 2365-2369.

Leydesdorff, L., & Opthof, T. (2011). Remaining problems with the "new crown indicator" (MNCS) of the CWTS. *Journal of Informetrics*, *5*(1), 224-225.

Leydesdorff, L., Radicchi, F., Bornmann, L., Castellano, C., & Nooy, W. (2013b). Field-normalized impact factors (IFs): A comparison of rescaling and fractionally counted IFs. *Journal of the American Society for Information Science and Technology*, *64*(11), 2299-2309.

Leydesdorff, L., Zhou, P., & Bornmann, L. (2013a). How can journal impact factors be normalized across fields of science? An assessment in terms of percentile ranks and fractional counts. *Journal of the American Society for Information Science and Technology*, *64*(1), 96-107.

Li, J., Sanderson, M., Willett, P., Norris, M., & Oppenheim, C. (2010). Ranking of library and information science researchers: Comparison of data sources for correlating citation data, and expert judgments. *Journal of Informetrics*, *4*(4), 554-563.

Li, Y., Radicchi, F., Castellano, C., & Ruiz-Castillo, J. (2013). Quantitative evaluation of alternative field normalization procedures. *Journal of Informetrics*, *7*(3), 746-755.

Lin, C. S., Huang, M. H., & Chen, D. Z. (2013). The influences of counting methods on university rankings based on paper count and citation count. *Journal of Informetrics*, *7*(3), 611-621.

Lindsey, D. (1980). Production and citation measures in the sociology of science: The problem of multiple authorship. *Social Studies of Science*, *10*(2), 145-162.

Lisée, C., Larivière, V., & Archambault, É. (2008). Conference proceedings as a source of scientific information: A bibliometric analysis. *Journal of the American Society for Information Science and Technology*, *59*(11), 1776-1784.

Liu, X. Z., & Fang, H. (2012a). Fairly sharing the credit of multi-authored papers and its application in the modification of h-index and g-index. *Scientometrics*, *91*(1), 37-49.

Liu, X. Z., & Fang, H. (2012b). Modifying h-index by allocating credit of multi-authored papers whose author names rank based on contribution. *Journal of Informetrics*, *6*(4), 557-565.

López-Illescas, C., De Moya-Anegón, F., & Moed, H. F. (2008). Coverage and citation impact of oncological journals in the Web of Science and Scopus. *Journal of Informetrics*, *2*(4), 304-316.

López-Illescas, C., De Moya Anegón, F., & Moed, H. F. (2009). Comparing bibliometric country-by-country rankings derived from the Web of Science and Scopus: The effect of poorly cited journals in oncology. *Journal of Information Science*, *35*(2), 244-256.

Lozano, G. A., Larivière, V., & Gingras, Y. (2012). The weakening relationship between the impact factor and papers' citations in the digital age. *Journal of the American Society for Information Science and Technology*, *63*(11), 2140-2145.

Lukovits, I., & Vinkler, P. (1995). Correct credit distribution: A model for sharing credit among coauthors. *Social Indicators Research*, *36*(1), 91-98.

Lundberg, J. (2007). Lifting the crown—Citation z-score. *Journal of informetrics*, *1*(2), 145-154.

Markpin, T., Boonradsamee, B., Ruksinsut, K., Yochai, W., Premkamolnetr, N., Ratchatahirun, P., & Sombatsompop, N. (2008). Article-count impact factor of materials science journals in SCI database. *Scientometrics*, *75*(2), 251-261.

Marshakova-Shaikevich, I. (1996). The standard impact factor as an evaluation tool of science fields and scientific journals. *Scientometrics*, *35*(2), 283-290.

Martin, B. R., & Irvine, J. (1983). Assessing basic research: Some partial indicators of scientific progress in radio astronomy. *Research Policy*, *12*(2), 61-90.

Marušić, A., Bošnjak, L., & Jerončić, A. (2011). A systematic review of research on the meaning, ethics and practices of authorship across scholarly disciplines. *PLOS ONE*, *6*(9), e23477.

Mayr, P., & Walter, A.-K. (2007). An exploratory study of Google Scholar. *Online Information Review*, *31*(6), 814-830.

Medoff, M. H. (2006). The efficiency of self-citations in economics. *Scientometrics*, *69*(1), 69-84.

Meho, L. I., & Rogers, Y. (2008). Citation counting, citation ranking, and h-index of human-computer interaction researchers: A comparison of Scopus and Web of Science. *Journal of the American Society for Information Science and Technology*, *59*(11), 1711-1726.

Meho, L. I., & Sugimoto, C. R. (2009). Assessing the scholarly impact of information studies: A tale of two citation databases—Scopus and Web of Science. *Journal of the American Society for Information Science and Technology*, *60*(12), 2499-2508.

Meho, L. I., & Yang, K. (2007). Impact of data sources on citation counts and rankings of LIS faculty: Web of Science versus Scopus and Google Scholar. *Journal of the American Society for Information Science and Technology*, *58*(13), 2105-2125.

Michels, C., & Fu, J. Y. (2014). Systematic analysis of coverage and usage of conference proceedings in Web of Science. *Scientometrics*, *100*(2), 307-327.

Michels, C., & Schmoch, U. (2012). The growth of science and database coverage. *Scientometrics*, *93*(3), 831-846.

Mikki, S. (2010). Comparing Google Scholar and ISI Web of Science for earth sciences. *Scientometrics*, *82*(2), 321-331.

Mingers, J. (2014). Problems with the SNIP indicator. *Journal of Informetrics*, *8*(4), 890-894.

Mingers, J., & Leydesdorff, L. (2015). A review of theory and practice in scientometrics. *European Journal of Operational Research*, *246*(1), 1-19.

Mingers, J., & Lipitakis, E. A. (2010). Counting the citations: A comparison of Web of Science and Google Scholar in the field of business and management. *Scientometrics*, *85*(2), 613-625.

Moed, H. F. (2002). Measuring China's research performance using the Science Citation Index. *Scientometrics*, *53*(3), 281-296.

Moed, H. F. (2005). *Citation analysis in research evaluation*. Springer.

Moed, H. F. (2010a). Measuring contextual citation impact of scientific journals. *Journal of Informetrics*, *4*(3), 265-277.

Moed, H. F. (2010b). CWTS crown indicator measures citation impact of a research group's publication oeuvre. *Journal of Informetrics*, *4*(3), 436-438.

Moed, H. F. (2016). Comprehensive indicator comparisons intelligible to non-experts: The case of two SNIP versions. *Scientometrics*, *106*(1), 51-65.

Moed, H. F., & Van Leeuwen, T. N. (1995). Improving the accuracy of Institute for Scientific Information's journal impact factors. *Journal of the American Society for Information Science*, *46*(6), 461-467.

Moed, H. F., & Van Leeuwen, T. N. (1996). Impact factors can mislead. *Nature*, *381*, 186.

Moed, H. F., Van Leeuwen, T. N., & Reedijk, J. (1998). A new classification system to describe the ageing of scientific journals and their impact factors. *Journal of Documentation*, *54*(4), 387-419.

Moed, H. F., Van Leeuwen, T. N., & Reedijk, J. (1999). Towards appropriate indicators of journal impact. *Scientometrics*, *46*(3), 575-589.

Mongeon, P., & Paul-Hus, A. (2016). The journal coverage of Web of Science and Scopus: A comparative analysis. *Scientometrics*, *106*(1), 213-228.

Moya-Anegón, F., Guerrero-Bote, V. P., Bornmann, L., & Moed, H. F. (2013). The research guarantors of scientific papers and the output counting: A promising new approach. *Scientometrics*, *97*(2), 421–434.

Mutz, R., & Daniel, H. D. (2012a). The generalized propensity score methodology for estimating unbiased journal impact factors. *Scientometrics*, *92*(2), 377-390.

Mutz, R., & Daniel, H. D. (2012b). Skewed citation distributions and bias factors: Solutions to two core problems with the journal impact factor. *Journal of Informetrics*, *6*(2), 169-176.

Nederhof, A. J. (2006). Bibliometric monitoring of research performance in the social sciences and the humanities: A review. *Scientometrics*, *66*(1), 81-100.

Nederhof, A. J., Van Leeuwen, T. N., & Clancy, P. (2012). Calibration of bibliometric indicators in space exploration research: A comparison of citation impact measurement of the space and ground-based life and physical sciences. *Research Evaluation*, *21*(1), 79-85.

Neuhaus, C., & Daniel, H. D. (2009). A new reference standard for citation analysis in chemistry and related fields based on the sections of Chemical Abstracts. *Scientometrics*, *78*(2), 219-229.

Nicolaisen, J. (2007). Citation analysis. *Annual Review of Information Science and Technology*, *41*, 609-641.

Nicolaisen, J., & Frandsen, T. F. (2008). The reference return ratio. *Journal of Informetrics*, *2*(2), 128-135.

Norris, M., & Oppenheim, C. (2007). Comparing alternatives to the Web of Science for coverage of the social sciences' literature. *Journal of Informetrics*, *1*(2), 161-169.

Norris, M., & Oppenheim, C. (2010). The h-index: A broad review of a new bibliometric indicator. *Journal of Documentation*, *66*(5), 681-705.

Olensky, M., Schmidt, M., & Van Eck, N. J. (in press). Evaluation of the citation matching algorithms of CWTS and iFQ in comparison to the Web of Science. *Journal of the Association for Information Science and Technology*.

Opthof, T., & Leydesdorff, L. (2010). Caveats for the journal and field normalizations in the CWTS ("Leiden") evaluations of research performance. *Journal of Informetrics*, *4*(3), 423-430.

Orduna-Malea, E., Ayllón, J. M., Martín-Martín, A., & López-Cózar, E. D. (2015). Methods for estimating the size of Google Scholar. *Scientometrics*, *104*(3), 931-949.

Ossenblok, T. L., Engels, T. C., & Sivertsen, G. (2012). The representation of the social sciences and humanities in the Web of Science—A comparison of publication patterns and incentive structures in Flanders and Norway (2005–9). *Research Evaluation*, *21*(4), 280-290.

Panaretos, J., & Malesios, C. (2009). Assessing scientific research performance and impact with single indices. *Scientometrics*, *81*(3), 635-670.

Perianes-Rodríguez, A., & Ruiz-Castillo, J. (2015). Multiplicative versus fractional counting methods for co-authored publications. The case of the 500 universities in the Leiden Ranking. *Journal of Informetrics*, *9*(4), 974-989.

Persson, O., Glänzel, W., & Danell, R. (2004). Inflationary bibliometric values: The role of scientific collaboration and the need for relative indicators in evaluative studies. *Scientometrics*, *60*(3), 421-432.

Pinski, G., & Narin, F. (1976). Citation influence for journal aggregates of scientific publications: Theory, with application to the literature of physics. *Information Processing and Management*, *12*(5), 297-312.

Plomp, R. (1990). The significance of the number of highly cited papers as an indicator of scientific prolificacy. *Scientometrics*, *19*(3), 185-197.

Plomp, R. (1994). The highly cited papers of professors as an indicator of a research group's scientific performance. *Scientometrics*, *29*(3), 377-393.

Pudovkin, A. I., & Garfield, E. (2004). Rank-normalized impact factor: A way to compare journal performance across subject categories. *Proceedings of the American Society for Information Science and Technology*, *41*(1), 507-515.

Pudovkin, A. I., & Garfield, E. (2009). Percentile rank and author superiority indexes for evaluating individual journal articles and the author's overall citation performance. *COLLNET Journal of Scientometrics and Information Management*,*3*(2), 3-10.

Radicchi, F., & Castellano, C. (2011). Rescaling citations of publications in physics. *Physical Review E*, *83*(4), 046116.

Radicchi, F., & Castellano, C. (2012a). Testing the fairness of citation indicators for comparison across scientific domains: The case of fractional citation counts. *Journal of Informetrics*, *6*(1), 121-130.

Radicchi, F., & Castellano, C. (2012b). A reverse engineering approach to the suppression of citation biases reveals universal properties of citation distributions. *PLOS ONE*, *7*(3), e33833.

Radicchi, F., Fortunato, S., & Castellano, C. (2008). Universality of citation distributions: Toward an objective measure of scientific impact. *Proceedings of the National Academy of Sciences of the United States of America*, *105*(45), 17268-17272.

Rinia, E. J., De Lange, C., & Moed, H. F. (1993). Measuring national output in physics: Delimitation problems. *Scientometrics*, *28*(1), 89-110.

Rons, N. (2012). Partition-based field normalization: An approach to highly specialized publication records. *Journal of Informetrics*, *6*(1), 1-10.

Rousseau, R. (2002). Journal evaluation: Technical and practical issues. *Library Trends*, *50*(3), 418-439.

Rousseau, R. (2005). Median and percentile impact factors: A set of new indicators. *Scientometrics*, *63*(3), 431-441.

Ruiz-Castillo, J. (2014). The comparison of classification-system-based normalization procedures with source normalization alternatives in Waltman and Van Eck (2013). *Journal of Informetrics*, *8*(1), 25-28.

Ruiz-Castillo, J., & Waltman, L. (2015). Field-normalized citation impact indicators using algorithmically constructed classification systems of science. *Journal of Informetrics*, *9*(1), 102-117.

Schreiber, M. (2007). Self-citation corrections for the Hirsch index. *EPL*, *78*(3), 30002.

Schreiber, M. (2008a). The influence of self-citation corrections on Egghe's g index. *Scientometrics*, *76*(1), 187-200.

Schreiber, M. (2008b). A modification of the h-index: The $h_m$-index accounts for multi-authored manuscripts. *Journal of Informetrics*, *2*(3), 211-216.

Schreiber, M. (2008c). To share the fame in a fair way, $h_m$ modifies h for multi-authored manuscripts. *New Journal of Physics*, *10*(4), 040201.

Schreiber, M. (2009a). A case study of the modified Hirsch index $h_m$ accounting for multiple coauthors. *Journal of the American Society for Information Science and Technology*, *60*(6), 1274-1282.

Schreiber, M. (2009b). Fractionalized counting of publications for the g-index. *Journal of the American Society for Information Science and Technology*, *60*(10), 2145-2150.

Schreiber, M. (2010a). How to modify the g-index for multi-authored manuscripts. *Journal of Informetrics*, *4*(1), 42-54.

Schreiber, M. (2010b). A case study of the modified g index: Counting multi-author publications fractionally. *Journal of Informetrics*, *4*(4), 636-643.

Schreiber, M. (2013). How much do different ways of calculating percentiles influence the derived performance indicators? A case study. *Scientometrics*, *97*(3), 821-829.

Schubert, A., & Braun, T. (1993). Reference standards for citation based assessments. *Scientometrics*, *26*(1), 21-35.

Schubert, A., & Braun, T. (1996). Cross-field normalization of scientometric indicators. *Scientometrics*, *36*(3), 311-324.

Schubert, A., Glänzel, W., & Braun, T. (1989). Scientometric datafiles. A comprehensive set of indicators on 2649 journals and 96 countries in all major science fields and subfields 1981–1985. *Scientometrics*, *16*(1), 3-478.

Schubert, A., Glänzel, W., & Thijs, B. (2006). The weight of author self-citations. A fractional approach to self-citation counting. *Scientometrics*, *67*(3), 503-514.

Seglen, P. O. (1992). The skewness of science. *Journal of the American Society for Information Science*, *43*(9), 628-638.

Seglen, P. O. (1994). Causal relationship between article citedness and journal impact. *Journal of the American Society for Information Science*, *45*(1), 1-11.

Seglen, P. O. (1997). Why the impact factor of journals should not be used for evaluating research. *British Medical Journal*, *314*, 497.

Sen, B. K. (1992). Normalised impact factor. *Journal of Documentation*, *48*(3), 318-325.

Shen, H. W., & Barabási, A. L. (2014). Collective credit allocation in science. *Proceedings of the National Academy of Sciences of the United States of America*, *111*(34), 12325-12330.

Sigogneau, A. (2000). An analysis of document types published in journals related to physics: Proceeding papers recorded in the Science Citation Index database. *Scientometrics*, *47*(3), 589-604.

Sirtes, D. (2012). Finding the Easter eggs hidden by oneself: Why Radicchi and Castellano's (2012) fairness test for citation indicators is not fair. *Journal of Informetrics*, *6*(3), 448-450.

Sivertsen, G., & Larsen, B. (2012). Comprehensive bibliographic coverage of the social sciences and humanities in a citation index: An empirical analysis of the potential. *Scientometrics*, *91*(2), 567-575.

Smolinsky, L. (2016). Expected number of citations and the crown indicator. *Journal of Informetrics*, *10*(1), 43-47.

Sombatsompop, N., & Markpin, T. (2005). Making an equality of ISI impact factors for different subject fields. *Journal of the American Society for Information Science and Technology*, *56*(7), 676-683.

Sombatsompop, N., Markpin, T., & Premkamolnetr, N. (2004). A modified method for calculating the impact factors of journals in ISI Journal Citation Reports: Polymer Science category in 1997–2001. *Scientometrics*, *60*(2), 217-235.

Stallings, J., Vance, E., Yang, J., Vannier, M. W., Liang, J., Pang, L., ... & Wang, G. (2013). Determining scientific impact using a collaboration index. *Proceedings of the National Academy of Sciences of the United States of America*, *110*(24), 9680-9685.

Stern, D. I. (2014). High-ranked social science journal articles can be identified from early citation information. *PLOS ONE*, *9*(11), e112520.

Thelwall, M. (2016). The precision of the arithmetic mean, geometric mean and percentiles for citation data: An experimental simulation modelling approach. *Journal of Informetrics*, *10*(1), 110-123.

Thelwall, M., & Fairclough, R. (2015). Geometric journal impact factors correcting for individual highly cited articles. *Journal of Informetrics*, *9*(2), 263-272.

Thelwall, M., & Kousha, K. (2015a). Web indicators for research evaluation, part 1: Citations and links to academic articles from the web. *El Profesional de la Información*, *24*(5), 587-606.

Thelwall, M., & Kousha, K. (2015b). Web indicators for research evaluation, part 2: Social media metrics. *El Profesional de la Información*, *24*(5), 607-620.

Thijs, B., & Glänzel, W. (2006). The influence of author self-citations on bibliometric meso-indicators. The case of European universities. *Scientometrics*, *66*(1), 71-80.

Tijssen, R. J., Visser, M. S., & Van Leeuwen, T. N. (2002). Benchmarking international scientific excellence: Are highly cited research papers an appropriate frame of reference? *Scientometrics*, *54*(3), 381-397.

Tol, R. S. (2011). Credit where credit's due: Accounting for co-authorship in citation counts. *Scientometrics*, *89*(1), 291-299.

Torres-Salinas, D., Lopez-Cózar, E. D., & Jiménez-Contreras, E. (2009). Ranking of departments and researchers within a university using two different databases: Web of Science versus Scopus. *Scientometrics*, *80*(3), 761-774.

Trueba, F. J., & Guerrero, H. (2004). A robust formula to credit authors for their publications. *Scientometrics*, *60*(2), 181-204.

Valderrama-Zurián, J.-C., Aguilar-Moya, R., Melero-Fuentes, D., & Aleixandre-Benavent, R. (2015). A systematic analysis of duplicate records in Scopus. *Journal of Informetrics*, *9*(3), 570-576.

Van Eck, N. J., & Waltman, L. (2014a). CitNetExplorer: A new software tool for analyzing and visualizing citation networks. *Journal of Informetrics*, *8*(4), 802-823.

Van Eck, N. J., & Waltman, L. (2014b). Visualizing bibliometric networks. In Y. Ding, R. Rousseau, & D. Wolfram (Eds.), *Measuring scholarly impact: Methods and practice* (pp. 285-320). Springer.

Van Eck, N. J., Waltman, L., Van Raan, A. F., Klautz, R. J., & Peul, W. C. (2013). Citation analysis may severely underestimate the impact of clinical research as compared to basic research. *PLOS ONE*, *8*(4), e62395.

Van Hooydonk, G. (1997). Fractional counting of multiauthored publications: Consequences for the impact of authors. *Journal of the American Society for Information Science*, *48*(10), 944-945.

Van Leeuwen, T. N., & Medina, C. C. (2012). Redefining the field of economics: Improving field normalization for the application of bibliometric techniques in the field of economics. *Research Evaluation*, *21*(1), 61-70.

Van Leeuwen, T. N., & Moed, H. F. (2002). Development and application of journal impact measures in the Dutch science system. *Scientometrics*, *53*(2), 249-266.

Van Leeuwen, T. N., & Moed, H. F. (2005). Characteristics of journal impact factors: the effects of uncitedness and citation distribution on the understanding of journal impact factors. *Scientometrics*, *63*(2), 357-371.

Van Leeuwen, T. N., Moed, H. F., Tijssen, R. J., Visser, M. S., & Van Raan, A. F. (2001). Language biases in the coverage of the Science Citation Index and its consequences for international comparisons of national research performance. *Scientometrics*, *51*(1), 335-346.

Van Leeuwen, T. N., Van der Wurff, L. J., & De Craen, A. J. M. (2007). Classification of "research letters" in general medical journals and its consequences in bibliometric research evaluation processes. *Research Evaluation*, *16*(1), 59-63.

Van Leeuwen, T. N., Visser, M. S., Moed, H. F., Nederhof, T. J., & Van Raan, A. F. (2003). The Holy Grail of science policy: Exploring and combining bibliometric tools in search of scientific excellence. *Scientometrics*, *57*(2), 257-280.

Van Leeuwen, T., Costas, R., Calero-Medina, C., & Visser, M. (2013). The role of editorial material in bibliometric research performance assessments. *Scientometrics*, *95*(2), 817-828.

Van Raan, A. F. J. (2004). Sleeping beauties in science. *Scientometrics*, *59*(3), 467-472.

Van Raan, A. F., Van Leeuwen, T. N., & Visser, M. S. (2011). Severe language effect in university rankings: Particularly Germany and France are wronged in citation-based rankings. *Scientometrics*, *88*(2), 495-498.

Van Raan, A. F., Van Leeuwen, T. N., Visser, M. S., Van Eck, N. J., & Waltman, L. (2010). Rivals for the crown: Reply to Opthof and Leydesdorff. *Journal of Informetrics*, *4*(3), 431-435.

Vanclay, J. K. (2012). Impact factor: Outdated artefact or stepping-stone to journal certification? *Scientometrics*, *92*(2), 211-238.

Vieira, E. S., & Gomes, J. A. (2009). A comparison of Scopus and Web of Science for a typical university. *Scientometrics*, *81*(2), 587-600.

Vieira, E. S., & Gomes, J. A. (2011). The journal relative impact: An indicator for journal assessment. *Scientometrics*, *89*(2), 631-651.

Vinkler, P. (2007). Eminence of scientists in the light of the h-index and other scientometric indicators. *Journal of Information Science*, *33*(4), 481-491.

Vinkler, P. (2010). *The evaluation of research by scientometric indicators*. Chandos Publishing.

Vinkler, P. (2012). The case of scientometricians with the "absolute relative" impact indicator. *Journal of Informetrics*, *6*(2), 254-264.

Visser, M. S., & Moed, H. F. (2008). *Comparing Web of Science and Scopus on a paper-by-paper basis*. Paper presented at the 10th International Conference on Science and Technology Indicators, Vienna, Austria.

Vrettas, G., & Sanderson, M. (2015). Conferences versus journals in computer science. *Journal of the Association for Information Science and Technology*, *66*(12), 2674-2684.

Wagner, C. S., & Leydesdorff, L. (2012). An integrated impact indicator: A new definition of 'impact' with policy relevance. *Research Evaluation*, *21*(3), 183-188.

Walters, W. H. (2007). Google Scholar coverage of a multidisciplinary field. *Information Processing and Management*, *43*(4), 1121-1132.

Walters, W. H. (2014). Do article influence scores overestimate the citation impact of social science journals in subfields that are related to higher-impact natural science disciplines? *Journal of Informetrics*, *8*(2), 421-430.

Waltman, L. (2012). An empirical analysis of the use of alphabetical authorship in scientific publishing. *Journal of Informetrics*, *6*(4), 700-711.

Waltman, L., Calero-Medina, C., Kosten, J., Noyons, E., Tijssen, R. J., Eck, N. J., ... & Wouters, P. (2012a). The Leiden Ranking 2011/2012: Data collection, indicators, and interpretation. *Journal of the American Society for Information Science and Technology*, *63*(12), 2419-2432.

Waltman, L., & Schreiber, M. (2013). On the calculation of percentile-based bibliometric indicators. *Journal of the American Society for Information Science and Technology*, *64*(2), 372-379.

Waltman, L., & Van Eck, N. J. (2012a). The inconsistency of the h-index. *Journal of the American Society for Information Science and Technology*, *63*(2), 406-415.

Waltman, L., & Van Eck, N. J. (2012b). A new methodology for constructing a publication-level classification system of science. *Journal of the American Society for Information Science and Technology*, *63*(12), 2378-2392.

Waltman, L., & Van Eck, N. J. (2013a). Source normalized indicators of citation impact: An overview of different approaches and an empirical comparison. *Scientometrics*, *96*(3), 699-716.

Waltman, L., & Van Eck, N. J. (2013b). A systematic empirical comparison of different approaches for normalizing citation impact indicators. *Journal of Informetrics*, *7*(4), 833-849.

Waltman, L., & Van Eck, N. J. (2015). Field-normalized citation impact indicators and the choice of an appropriate counting method. *Journal of Informetrics*, *9*(4), 872-894.

Waltman, L., Van Eck, N. J., Van Leeuwen, T. N., & Visser, M. S. (2013). Some modifications to the SNIP journal impact indicator. *Journal of informetrics*, *7*(2), 272-285.

Waltman, L., Van Eck, N. J., Van Leeuwen, T. N., Visser, M. S., & Van Raan, A. F. (2011a). Towards a new crown indicator: Some theoretical considerations. *Journal of Informetrics*, *5*(1), 37-47.

Waltman, L., Van Eck, N. J., Van Leeuwen, T. N., Visser, M. S., & Van Raan, A. F. (2011b). Towards a new crown indicator: An empirical analysis. *Scientometrics*, *87*(3), 467-481.

Waltman, L., Van Eck, N. J., & Van Raan, A. F. (2012b). Universality of citation distributions revisited. *Journal of the American Society for Information Science and Technology*, *63*(1), 72-77.

Waltman, L., & Yan, E. (2014). PageRank-related methods for analyzing citation networks. In Y. Ding, R. Rousseau, & D. Wolfram (Eds.), *Measuring scholarly impact: Methods and practice* (pp. 83-100). Springer.

Wan, J. K., Hua, P. H., & Rousseau, R. (2007). The pure h-index: Calculating an author's h-index by taking co-authors into account. *COLLNET Journal of Scientometrics and Information Management*, *1*(2), 1-5.

Wan, X., & Liu, F. (2014). WL-index: Leveraging citation mention number to quantify an individual's scientific impact. *Journal of the Association for Information Science and Technology*, *65*(12), 2509-2517.

Wang, J. (2013). Citation time window choice for research impact evaluation. *Scientometrics*, *94*(3), 851-872.

Wang, Q., & Waltman, L. (2015). *Large-scale analysis of the accuracy of the journal classification systems of Web of Science and Scopus*. arXiv:1511.00735.

West, J. D., Bergstrom, T. C., & Bergstrom, C. T. (2010a). The eigenfactor metrics: A network approach to assessing scholarly journals. *College and Research Libraries*, *71*(3), 236-244.

West, J., Bergstrom, T., & Bergstrom, C. T. (2010b). Big Macs and eigenfactor scores: Don't let correlation coefficients fool you. *Journal of the American Society for Information Science and Technology*, *61*(9), 1800-1807.

Wildgaard, L. (2015). A comparison of 17 author-level bibliometric indicators for researchers in astronomy, environmental science, philosophy and public health in Web of Science and Google Scholar. *Scientometrics*, *104*(3), 873-906.

Wildgaard, L., Schneider, J. W., & Larsen, B. (2014). A review of the characteristics of 108 author-level bibliometric indicators. *Scientometrics*, *101*(1), 125-158.

Wouters, P., & Costas, R. (2012). *Users, narcissism and control: Tracking the impact of scholarly publications in the 21st century*. SURFfoundation.

Wouters, P., Thelwall, M., Kousha, K., Waltman, L., De Rijcke, S., Rushforth, A., & Franssen, T. (2015). *The metric tide: Literature review (Supplementary report I to* *the independent review of the role of metrics in research assessment and management)*. HEFCE.

Wuchty, S., Jones, B. F., & Uzzi, B. (2007). The increasing dominance of teams in production of knowledge. *Science*, *316*, 1036-1039.

Xu, J., Ding, Y., Song, M., & Chambers, T. (in press). Author credit-assignment schemas: A comparison and analysis. *Journal of the Association for Information Science and Technology*.

Zhang, J., Yu, Q., Zheng, F., Long, C., Lu, Z., & Duan, Z. (in press). Comparing keywords plus of WOS and author keywords: A case study of patient adherence research. *Journal of the Association for Information Science and Technology*.

Zhang, L., & Glänzel, W. (2012). Proceeding papers in journals versus the "regular" journal publications. *Journal of Informetrics*, *6*(1), 88-96.

Zhang, Z., Cheng, Y., & Liu, N. C. (2014). Comparison of the effect of mean-based method and z-score for field normalization of citations at the level of Web of Science subject categories. *Scientometrics*, *101*(3), 1679-1693.

Zhu, X., Turney, P., Lemire, D., & Vellino, A. (2015). Measuring academic influence: Not all citations are equal. *Journal of the Association for Information Science and Technology*, *66*(2), 408-427.

Zitt, M. (2010). Citing-side normalization of journal impact: A robust variant of the audience factor. *Journal of Informetrics*, *4*(3), 392-406.

Zitt, M., Ramanana-Rahary, S., & Bassecoulard, E. (2003). Correcting glasses help fair comparisons in international science landscape: Country indicators as a function of ISI database delineation. *Scientometrics*, *56*(2), 259-282.

Zitt, M., Ramanana-Rahary, S., & Bassecoulard, E. (2005). Relativity of citation performance and excellence measures: From cross-field to cross-scale effects of field-normalisation. *Scientometrics*, *63*(2), 373-401.

Zitt, M., & Small, H. (2008). Modifying the journal impact factor by fractional citation weighting: The audience factor. *Journal of the American Society for Information Science and Technology*, *59*(11), 1856-1860.